\documentclass[pra,twocolumn,showpacs,superscriptaddress]{revtex4}

\usepackage{graphicx}
\usepackage{dcolumn}
\usepackage{bm}

\begin{document}

\preprint{APS/123-QED}

\title{Coherence Protection by the Quantum Zeno Effect and Non-Holonomic Control\\in a Rydberg Rubidium isotope}

\author{E.~Brion}
\affiliation{Laboratoire Aim\'{e} Cotton, CNRS II, B\^atiment 505, 91405 Orsay Cedex, France.}

\author{V.~M.~Akulin}
\affiliation{Laboratoire Aim\'{e} Cotton, CNRS II, B\^atiment 505, 91405 Orsay Cedex, France.}

\author{D.~Comparat}
\affiliation{Laboratoire Aim\'{e} Cotton, CNRS II, B\^atiment 505, 91405 Orsay Cedex, France.}

\author{I.~Dumer}
\affiliation{College of Engineering, University of California, Riverside, CA 92521, USA.}

\author{G.~Harel}
\affiliation{Department of Computing, University of Bradford, Bradford, West Yorkshire BD7 1DP, United Kingdom.}

\author{N.~K\'{e}baili}
\affiliation{Laboratoire Aim\'{e} Cotton, CNRS II, B\^atiment 505, 91405 Orsay Cedex, France.}

\author{G.~Kurizki}
\affiliation{Department of Chemical Physics, Weizmann Institue of Science, 76100 Rehovot, Israel.}

\author{I.~Mazets}
\affiliation{Department of Chemical Physics, Weizmann Institue of Science, 76100 Rehovot, Israel.}
\affiliation{A.F. Ioffe Physico-Technical Institute, 194021 St. Petersburg, Russia.}

\author{P.~Pillet}
\affiliation{Laboratoire Aim\'{e} Cotton, CNRS II, B\^atiment 505, 91405 Orsay Cedex, France.}

\date{\today}

\widetext


\begin{abstract}
The protection of the coherence of open quantum systems against the influence
of their environment is a very topical issue. A scheme is proposed here which
protects a general quantum system from the action of a set of arbitrary
uncontrolled unitary evolutions. This method draws its inspiration from ideas 
of standard error-correction (ancilla adding, coding and decoding)
and the Quantum Zeno Effect. A demonstration of our method on a
simple atomic system, namely a Rubidium isotope, is proposed.
\end{abstract}

\maketitle

\section{Introduction\label{I}}

The uncontrollable interaction of an open quantum system with its environment
leads to complete loss of the information initially stored in its quantum
state. This phenomenon is commonly referred to as ''loss of coherence''. The
question of how it is possible to avoid the negative influence of this process
is one of the most interesting issues in modern quantum mechanics, and
concerns many different fields of physics, in particular the domains of quantum
information and computation.

In the context of quantum information, the effects of interactions with the
environment, known as ''quantum errors'', may render information
storage and processing unreliable \cite{Chuang,Preskill}. Since
Shor's demonstration that error-correcting schemes exist in quantum computation \cite{SHOR},
 a general framework of error-correction
has been built upon the formalism of quantum operations. The main
contributions concern quantum codes \cite{Knill1}, and particularly the
class of stabilizer codes \cite{Gottesman1,Gottesman2}; other
strategies developed suggest the use of ''noiseless quantum
codes'' or ''decoherence-free subspaces'' \cite{Zanardi1,Knill2,Lidar}. All
these methods usually demand that errors act independently on different qubits
(the independent error model), and make use of the symmetry properties associated
with these requirements. This implies that the set of errors to be corrected
hence is restricted to a special subgroup, called the Clifford group. In this
paper, we present a protection method which draws its inspiration from the ideas 
of standard error-correction and the Quantum Zeno Effect, and
requires no specific symmetry of the errors. Moreover, we suggest its
physical implementation in an arbitrary quantum system, and show how it
works for the example of a Rubidium isotope.

The phenomenon known as the Quantum Zeno Effect (QZE) takes place in a system which
is subject to frequent measurements projecting it onto its (necessarily
known) initial state: if the time interval between two projections is small
enough the evolution of the system is nearly ''frozen''. This effect, and its inverse (the anti-Zeno effect),
have been widely investigated theoretically \cite{TQZE1,TQZE2,AQZE1,AQZE2} as
well as experimentally \cite{EQZE1,EQZE2}. Generalizations have been
proposed which employ incomplete measurements \cite{GQZE1}: in this
setting, the Hilbert space is split into ''Zeno subspaces'' (degenerate 
multidimensional eigenspaces of the measured
observable), and the state vector of the system is compelled by frequent
measurements of the physical observable to remain in its initial Zeno subspace. 
The dynamics of the system in the Zeno subspaces has also been studied in different specific situations
\cite{GQZE2}.

Employing these ideas, enriched by standard techniques from coding theory
\cite{GALLAGER}, we have previously proposed an information protection scheme \cite{CPZE}
in Zanardi's spirit \cite{Zanardi2}, except that we do not make
any symmetry assumption on the unitary errors we consider. We form a compound
system $\mathcal{S}$ which comprises the information system $\mathcal{I}$ to
be protected and an auxiliary system $\mathcal{A}$ (called ancilla). We then
apply a controlled unitary operation $\widehat{C}$ (the coding
matrix) which encodes the information, initially stored in $\mathcal{I}$, in
an entangled state of $\mathcal{I}$ and $\mathcal{A}$. After a short time
interval, during which infinitesimal errors may have occurred, we apply the
unitary transformation $\widehat{C}^{-1}$ (the inverse to the previous step),
which decodes information. Finally, we measure the ancilla to get rid of the
infinitesimal changes that may have been caused by errors. Whereas in classical
error-correction theory, the ancilla contains information about the errors allowing 
them to be corrected, in
our QZE-based approach, the quantum state of
the ancilla resulting from an elementary (coding-errors-decoding) sequence
is close to its initial state, so that the measurement of the
ancilla brings it back to its initial state with a probability of nearly
1. The key point of our method is to entangle the initial state of the ancilla
with the state of the system in such a way that the detection of the ancilla
in its initial state implies that the system is also in its initial state.
This is achieved through the coding procedure which has to ensure that after
the exposition of the system to the action of errors, the initial state of the
ancilla remains entangled only with the initial state of the system, whereas
any other entanglement remains small during the time interval between consecutive 
measurements.

The coding procedure involves a rather complex unitary transformation in the
Hilbert space of the compound system: performing such coding operations is a
non-trivial quantum control problem in itself. However, one can achieve this
objective by adapting the idea of Non-Holonomic Control which we have
previously presented \cite{NHC}. Specific algorithms which allow us to determine and physically implement the
coding matrix $\widehat{C}$ have been constructed.

In this paper, we provide a comprehensive presentation of our theoretical scheme, including algorithmic aspects which were not dealt with in \cite{CPZE}. Moreover, we propose a ''demonstrative'' application of our technique to a physical system : more precisely, we show how our method can protect one qubit of information stored in the spin variable of the quantum state of a single Rubidium atom, the orbital variable
playing the role of the ancilla, against a given set of error inducing Hamiltonians. Here we describe a realistic experimental
setting which achieves the different steps of our scheme through the
application of a sequence of laser pulses and culminates in a measurement 
involving spontaneous emission. When dealing with ensembles of atoms, as usually done in current cold atom experiments, experimental drawbacks arise due to dipolar interactions which forbid the actual implementation of our application. Yet, even though not completely satisfactory from an experimental point of view, the example we propose here shows the general scope of our method as well as its physical operationality.

The paper is organized as follows. In Sec.II, we present a
multidimensional generalization of the QZE and its application
to the protection of information contained in compound systems. In Sec.III, 
we present the algorithms which enable us to calculate the code space
and physically implement the coding matrix through the non holonomic
control technique. In Sec.IV, we focus on the application of our
method to a Rubidium isotope. In Appendix A, we present the explicit 
derivation of the code subspace.

\section{Multidimensional Zeno Effect and Coherence Protection\label{II}}

We start this section by the geometric presentation of a multidimensional
QZE which allows us to protect an arbitrary subspace of the
Hilbert space against the action of a set of given interaction Hamiltonians.
In the second part of this section, we take advantage of this phenomenon to
protect an information-carrying subsystem of a compound quantum system from
the influence of some uncontrolled error-inducing external fields.

\bigskip

Consider a quantum system $\mathcal{S}$, whose $N$-dimensional Hilbert space
is denoted by $\mathcal{H}$ and whose time-dependent Hamiltonian has the
form
\begin{equation}
\widehat{H}(\tau)=\sum_{m=1}^{M}f_{m}(\tau)\widehat{E}_{m},
\label{Hamiltonian}%
\end{equation}
where $\left\{  \widehat{E}_{m}\right\}  _{m=1,...,M}$ are $M$ given
independent Hermitian matrices on $\mathcal{H}$ and $\left\{  f_{m}%
(\tau)\right\}  _{m=1,...,M}$\ are $M$ unknown functions of time. The Hamiltonian $\widehat
{H}(\tau)$ accounts for the errors we want to get rid of. Note that the unperturbed part of the Hamiltonian (\ref{Hamiltonian}) is assumed to be zero (or proportional to the identity so that one can set it to zero). The standard
QZE \cite{TQZE1,TQZE2,AQZE1,AQZE2} implies that we can nearly ''freeze'' the
evolution of the system by measuring it frequently enough in its (known) initial
state ; in other words, this effect allows us to protect the
one-dimensional subspace spanned by the initial state of the system from the
influence of the error-inducing Hamiltonian (\ref{Hamiltonian}). In what
follows, we generalize this effect so as to protect an arbitrary
multidimensional subspace $\mathcal{C}$ from $\widehat{H}(\tau)$.

Any vector $\left|  \psi\right\rangle $ of $\mathcal{C}$ evolves according to
the operator
\[
\widehat{U}(t,t_{0})=\mathcal{T}\left\{  \exp\left[  -i\int_{t_{0}}%
^{t}\widehat{H}(\tau)d\tau\right]  \right\}  ,
\]
where $\mathcal{T}$ denotes time-ordering, and where we set
$\hbar=1$. For the QZE to hold, we shall only
consider evolution in short time periods, whose duration $T$ is so short that the corresponding 
action of the $M$ components of the Hamiltonian (\ref{Hamiltonian}) is small, i.e. $\left|  \widehat{E}_{m}%
\int_{t}^{t+T}f_{m}(\tau)d\tau\right|  \ll1.$ We can thus expand

\begin{equation}
\widehat{U}(t+T,t)=\widehat{U}_{inf}\simeq\widehat{I}-i\sum_{m=1}^{M}\left(
\int_{t}^{t+T} f_{m}(\tau)d\tau\right)  \widehat{E}_{m}. \label{evolinf}%
\end{equation}
This implies that after a Zeno interval $T$, the initial state $\left|
\psi\right\rangle $ is transformed into $\left|  \psi_{e}\right\rangle
=\left|  \psi\right\rangle +\left|  \delta\psi_{e}\right\rangle $ where
$\left|  \delta\psi_{e}\right\rangle \simeq-i\sum_{m=1}^{M}\varepsilon
_{m}\widehat{E}_{m}\left|  \psi\right\rangle $ with $\varepsilon_{m}=\left(
\int f_{m}(\tau)d\tau\right)  .$ Note that, strictly speaking, the operator $\widehat{U}_{inf}$ introduced in Eq.(\ref{evolinf}) is not unitary : nevertheless, the non-unitary part, due to the truncation of the time-development of the evolution operator, is of second order in time and is thus negligible in the Zeno limit $\left( T \rightarrow 0 \right)$. Moreover, as we exclusively consider finite dimensional systems interacting with classical external fields, the approximation Eq.(2) holds, without raising any mathematical problem. But it is worth emphasizing that this is no longer the case when dealing with systems of infinitely large Hilbert space (for example, see \cite{RBMB04}).

\begin{figure*}
[floatfix]
\begin{center}
\includegraphics[
height=6.5363in,
width=4.5956in
]%
{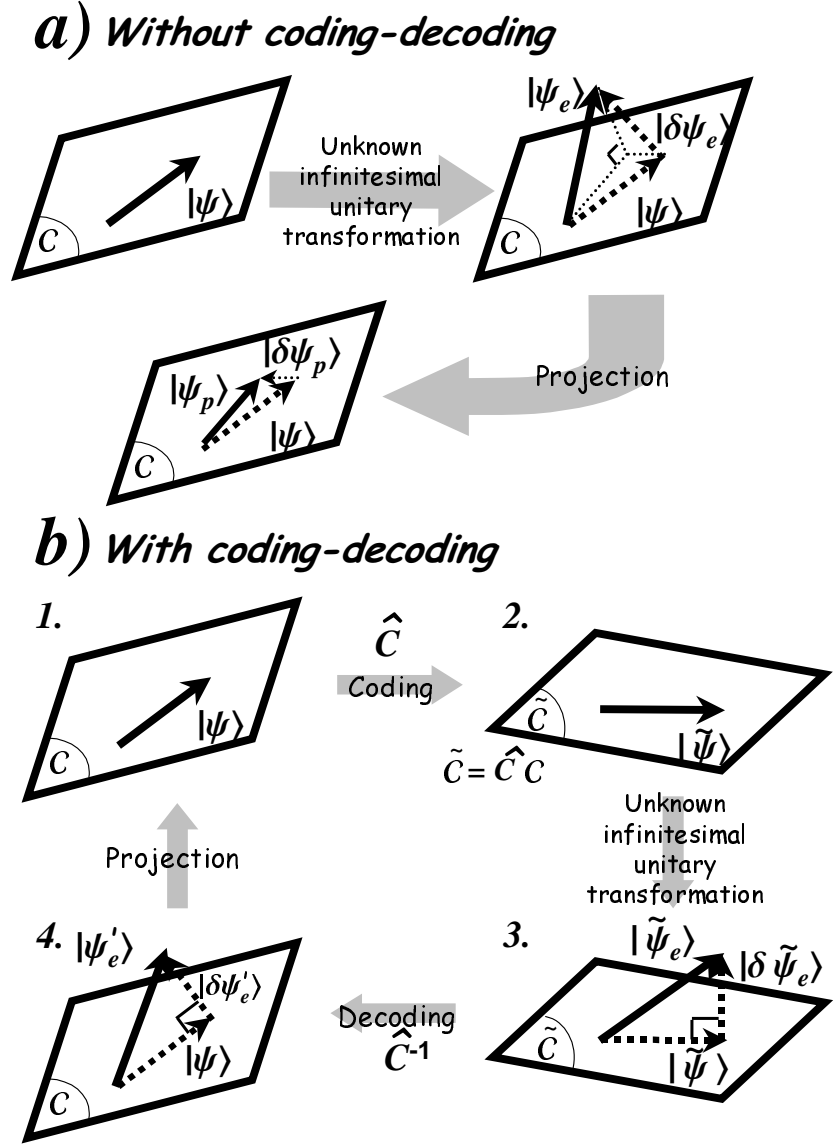}%
\caption{Multidimensional QZE: a) a simple
projection fails to recover the initial vector, b) the sequence
coding-decoding-projection protects the initial vector.}%
\label{fig1}%
\end{center}
\end{figure*}

Let us assume that we are physically able to perform a measurement-induced
projection onto $\mathcal{C}$ in the system $\mathcal{S}$ (see below the 
discussion of such projections for compound systems comprising an information 
subsystem and an ancilla). Now if we just
follow the standard QZE procedure and merely
project the state vector $\left|  \psi_{e}\right\rangle ,$ resulting from the
infinitesimal evolution of the initial state $\left|  \psi\right\rangle ,$
onto $\mathcal{C}$, we obtain a vector $\left|  \psi_{p}\right\rangle $, which
(a priori) differs from $\left|  \psi\right\rangle $ (see Fig.\ref{fig1}a).
This is due to the fact that usually the operators $\left\{  \widehat{E}%
_{m}\right\}  _{m=1,...,M}$\ do not act orthogonally on $\mathcal{C}$, which
means that the vectors $\widehat{E}_{m}\left|  \psi\right\rangle $\ and thus
the increment vector $\left|  \delta\psi_{e}\right\rangle $\ itself are not
orthogonal to $\mathcal{C}$. It is a well-known manifestation of the standard Zeno Effect : the system is compelled to remain in a ''Zeno subspace'' (which corresponds here to $\mathcal{C}$) in which it presents a remaining dynamics called ''Zeno dynamics'' \cite{GQZE2} ; in our case, this dynamics is unknown and thus threatens the information stored in $\mathcal{C}$. Therefore, we see that the standard Zeno strategy does not suffice to protect a multidimensional subspace : we have to adapt it using some ideas of coding theory.

To this end, we assume a unitary matrix $\widehat{C}$ acting on
$\mathcal{H}$, which we call the coding matrix, such that the Hermitian
operators $\left\{  \widehat{E}_{m}\right\}  _{m=1,...,M}$ act orthogonally on
the subspace $\widetilde{\mathcal{C}}=\widehat{C}\mathcal{C}$, which we call
the code space. Let us denote by $I\geq1$ the dimension of $\mathcal{C}$ and by $\left\{  \left|
\gamma_{i}\right\rangle \right\}  _{i=1,...,I}$ one of its orthonormal bases ; $\left\{  \left|  \widetilde{\gamma}%
_{i}\right\rangle =\widehat{C}\left|  \gamma_{i}\right\rangle \right\}
_{i=1,...,I}$ will denote one of the orthonormal bases of $\widetilde{\mathcal{C}}$, the state vectors $\left|  \widetilde{\gamma}
_{i}\right\rangle$ being called the codewords. For
any pair $\left(  \left|  \widetilde{\gamma}_{s}\right\rangle ,\left|
\widetilde{\gamma}_{t}\right\rangle \right)  $ of codewords\ and any operator
$\widehat{E}_{m}\in\left\{  \widehat{E}_{m}\right\}  _{m=1,...,M}$ we have, by
the definitions of $\widehat{C}$ and $\widetilde{\mathcal{C}}$
\begin{eqnarray}
\left\langle \widetilde{\gamma}_{t}|\widetilde{\gamma}_{s}\right\rangle=\delta_{st}\mbox{(orthonormality condition),}\label{bascod1}\\ \left\langle \widetilde{\gamma}_{t}\left|  \widehat{E}_{m}\right|  \widetilde{\gamma}_{s}\right\rangle=0\mbox{(orthogonality of the errors).} \label{bascod2}%
\end{eqnarray}
Equivalently, for any pair $\left(  \left|  \psi\right\rangle ,\left|
\chi\right\rangle \right)$ of vectors of $\mathcal{C}$ and for any operator
$\widehat{E}_{m}\in\left\{  \widehat{E}_{m}\right\}  _{m=1,...,M}$
\begin{equation}
\left\langle \chi\left|  \widehat{C}^{\dagger}\widehat{E}_{m}\widehat
{C}\right|  \psi\right\rangle =0. \label{matcod}%
\end{equation}
In particular, for any pair $\left(  \left|  \gamma_{s}\right\rangle
,\left|  \gamma_{t}\right\rangle \right)  $ of basis vectors of $\mathcal{C}$
and for any operator $\widehat{E}_{m}\in\left\{  \widehat{E}_{m}\right\}
_{m=1,...,M}$
\begin{equation}
\left\langle \gamma_{t}\left|  \widehat{C}^{\dagger}\widehat{E}_{m}\widehat
{C}\right|  \gamma_{s}\right\rangle =0. \label{matcodbis}%
\end{equation}
If we apply the coding matrix $\widehat{C}$\ to the
initial state vector $\left|  \psi\right\rangle $, before exposing it to the
action of the Hamiltonian (\ref{Hamiltonian}), we obtain the new vector
$\left|  \widetilde{\psi}\right\rangle =\widehat{C}\left|  \psi\right\rangle
\in\widetilde{\mathcal{C}}$ (Fig.\ref{fig1}b1,2) which is transformed after a
Zeno interval $T$ into $\left|  \widetilde{\psi}_{e}\right\rangle =\widehat
{U}_{inf}\left|  \widetilde{\psi}\right\rangle =\left|  \widetilde{\psi
}\right\rangle +\left|  \delta\widetilde{\psi}_{e}\right\rangle ,$ where
$\left|  \delta\widetilde{\psi}_{e}\right\rangle \simeq-i\sum_{m=1}%
^{M}\varepsilon_{m}\widehat{E}_{m}\left|  \widetilde{\psi}\right\rangle
=-i\sum_{m=1}^{M}\varepsilon_{m}\widehat{E}_{m}\widehat{C}\left|
\psi\right\rangle $ (Fig.\ref{fig1}b3).\ Decoding $\left|  \widetilde{\psi
}_{e}\right\rangle $\ yields the vector $\left|  \psi_{e}^{\prime
}\right\rangle =\widehat{C}^{-1}\left|  \widetilde{\psi}_{e}\right\rangle
=\left|  \psi\right\rangle +\left|  \delta\psi_{e}^{\prime}\right\rangle $
where $\left|  \delta\psi_{e}^{\prime}\right\rangle \simeq-i\sum_{m=1}%
^{M}\varepsilon_{m}\widehat{C}^{\dagger}\widehat{E}_{m}\widehat{C}\left|
\psi\right\rangle $. From Eq.(\ref{matcod}) it can be seen that for any vector
$\left|  \chi\right\rangle \in\mathcal{C}$, $\left\langle \chi|\delta\psi
_{e}^{\prime}\right\rangle =-i\sum_{m=1}^{M}\varepsilon_{m}\left\langle
\chi\right|  \widehat{C}^{\dagger}\widehat{E}_{m}\widehat{C}\left|
\psi\right\rangle =0$ which means that $\left|  \delta\psi_{e}^{\prime
}\right\rangle $ is orthogonal to $\mathcal{C}$ (Fig.\ref{fig1}b4).
A measurement-induced projection onto $\mathcal{C}$\ finally recovers the
initial vector $\left|  \psi\right\rangle $ with a probability very close to
$1$ (the error probability is proportional to $T^{2}$). If the (coding-decoding-projection) sequence 
is frequently repeated, any vector $\left|
\psi\right\rangle $ of the subspace $\mathcal{C}$\ can thus be protected from
the Hamiltonian (\ref{Hamiltonian}) for as long as needed. We stress that the role of projective measurements consists both in confining the system in $\mathcal{C}$ (as in the standard Quantum Zeno Effect) and in clearing out the erroneous component which has been made orthogonal to $\mathcal{C}$ through coding and decoding.

Let us note that a more general version of conditions (\ref{bascod2}) can
be considered. Indeed, if for any pair of
codewords $\left(  \left|  \widetilde{\gamma}_{s}\right\rangle ,\left|
\widetilde{\gamma}_{t}\right\rangle \right)  $ of $\widetilde{\mathcal{C}}$\ and any error Hamiltonian $\widehat{E}_{m}\in\left\{  \widehat{E}_{m}\right\},$
\[
\left\langle \widetilde{\gamma}_{t}\left|  \widehat{E}_{m}\right|
\widetilde{\gamma}_{s}\right\rangle =\delta_{ts}\xi_{m},
\]
where $\delta_{ts}$ is the Kronecker symbol and $\xi_{m}$ a real number
depending only on the number $m$ of the error Hamiltonian $\widehat{E}_{m}$,
the projection onto
$\mathcal{C}$ $=Span\left\{  \left|  \gamma_{i}\right\rangle
,i=1,...,I\right\}  $\ of the state vector $\left|  \psi_{e}^{\prime
}\right\rangle =\left|  \psi\right\rangle -i\sum_{m=1}^{M}\varepsilon
_{m}\widehat{E}_{m}\left|  \psi\right\rangle $, obtained after a
(coding-decoding) sequence, yields 
\[
\widehat{\Pi}_{C}\left|  \psi_{e}^{\prime
}\right\rangle =\left|  \psi\right\rangle -i\sum_{m=1}^{M}\varepsilon
_{m}\widehat{\Pi}_{C}\widehat{E}_{m}\left|  \psi\right\rangle
\]
where
$\widehat{\Pi}_{C}=\left[  \sum_{t=1}^{I}\left|  \gamma_{t}\right\rangle
\left\langle \gamma_{t}\right|  \right]  $ ; if we denote by $\left|
\psi\right\rangle =\sum_{s=1}^{I}\alpha_{s}\left|  \gamma_{s}\right\rangle $
the decomposition of the initial information state vector, $\widehat{\Pi}%
_{C}\widehat{E}_{m}\left|  \psi\right\rangle $ has the form
\begin{eqnarray}
\widehat{\Pi}_{C}\widehat{E}_{m}\left|  \psi\right\rangle & = & \sum
_{s,t=1}^{I}\alpha_{s}\left|  \gamma_{t}\right\rangle \left\langle \gamma
_{t}\left|  \widehat{C}^{\dagger}\widehat{E}_{m}\widehat{C}\right|  \gamma
_{s}\right\rangle  \nonumber \\
& = & \xi_{m}\sum_{s,t=1}^{I}\alpha_{s}\delta_{ts}\left|  \gamma_{t}%
\right\rangle  \nonumber \\
& = &\xi_{m}\left|  \psi\right\rangle , \nonumber 
\end{eqnarray}
which finally leads to $\widehat{\Pi}_{C}\left|  \psi_{e}^{\prime
}\right\rangle =\left(  1-i\sum_{m=1}^{M}\varepsilon_{m}\xi_{m}\right)
\left|  \psi\right\rangle $. In other words, the errors $\widehat{E}_{m}$ just
introduce a global phase factor in front of the initial information state
vector, but leaves its coherence intact.
Obviously, the correction conditions (\ref{bascod2}) are obtained as a particular
case of the above conditions, setting $\xi_{m}=0$ for all $m$.
Yet, though less general, they will be employed in the rest of the paper for the 
sake of simplicity.

The multidimensional generalization of the QZE we have just
described allows us to protect any subspace $\mathcal{C}$ of a Hilbert space
$\mathcal{H}$\ against Hamiltonians of the form (\ref{Hamiltonian}) provided the projection onto $\mathcal{C}$\ is physically achievable and
the coding matrix $\widehat{C}$ exists. This result is very useful in the context of information protection
as we will show in the following paragraphs.

\bigskip

Consider an information system $\mathcal{I}$ of Hilbert space $\mathcal{H}%
_{I}$ and dimensionality $I$. This system is subjected to a set of $M$
error-inducing Hamiltonians $\left\{  \widehat{E}_{m}\right\}  _{m=1,...,M}$
which, for instance, represent interactions of the system with $M$
uncontrolled external fields $f_{m}(t)$: we want to get rid of this external
influence which is likely to result in the loss of the information stored in
the initial state vector $\left|  \psi_{I}\right\rangle =\sum_{i=1}^{I}%
c_{i}\left|  \nu_{i}\right\rangle $, where $\left\{  \left|  \nu
_{i}\right\rangle \right\}  _{i=1,...,I}$ denotes an orthonormal basis of
$\mathcal{H}_{I}.$ To this end, we will use the multidimensional Zeno Effect.
As the multidimensional QZE can only protect a subspace of the
whole Hilbert space, we first have to add an $A$-dimensional auxiliary system
$\mathcal{A}$\ (called ancilla) to our system $\mathcal{I}$, so that the
information is transferred from $\mathcal{H}_{I}$\ into an $I$-dimensional
subspace $\mathcal{C}$ of the $\left(  N=I\times A\right)  $-dimensional
Hilbert space $\mathcal{H}=\mathcal{H}_{I}\mathcal{\otimes H}_{A}$\ of the
compound system $\mathcal{S=I\otimes A}$ (let us note that this ancilla adding procedure is quite standard in quantum error-correction \cite{Chuang} and corresponds to the redundancy used in classical coding methods). Furthermore, we shall suppose that all the state
vectors of the different Hilbert spaces $\mathcal{H}_{I}$, $\mathcal{H}_{A}$
and hence $\mathcal{H}$\ are degenerate in energy so that the unperturbed part $\widehat
{H}_{0}$ of the Hamiltonian can be set to zero as in the first part of this section: the subspace $\mathcal{C}$ and the information it carries can thus be protected through the multidimensional
QZE (in Sec.IV, on the example of Rb, we shall see that the multidimensional QZE may also be used even though $\widehat
{H}_{0}$ is not zero, provided $\widehat {H}_{0}$ and the errors have some convenient properties). Note that $\mathcal{A}$ and $\mathcal{I}$ need not be ''physically separate'' systems, but only have to possess independent Hilbert spaces $\mathcal{H}_{A}$ and $\mathcal{H}_{I}$. For example, in Sec.IV, we shall consider the Rubidium atom as the compound of two independent subsystems, namely its spin (which plays the role of $\mathcal{I}$) and orbital part (which plays the role of $\mathcal{A}$). Doing so, we shall use the terms 'factorized' and 'entangled' in a generalized manner to designate states obtained as a direct product of the spin and the orbital parts, and linear combinations of such states, respectively.

Let us now return to our problem and first consider the simple case in which the ancilla
is initially in the pure state $\left|  \alpha\right\rangle $. The information
previously carried by $\left|  \psi_{I}\right\rangle \in\mathcal{H}_{I}$ is
then transferred into the factorized state $\left|  \psi\right\rangle =\left|
\psi_{I}\right\rangle \otimes\left|  \alpha\right\rangle =$ $\sum_{i=1}%
^{I}c_{i}\left|  \nu_{i}\right\rangle \otimes\left|  \alpha\right\rangle =$
$\sum_{i=1}^{I}c_{i}\left|  \gamma_{i}\right\rangle $ which belongs to the
tensor product subspace $\mathcal{C}=\mathcal{H}_{I}\mathcal{\otimes
}Span\left[  \left|  \alpha\right\rangle \right]  =Span\left[  \left\{
\left|  \gamma_{i}\right\rangle =\left|  \nu_{i}\right\rangle \otimes\left|
\alpha\right\rangle \right\}  _{i=1,...,I}\right]  $. In other words, the
initial density matrix of the compound system $\mathcal{S}$ is
$\widehat{\rho}=\left(  \left|  \psi_{I}\right\rangle \left\langle \psi
_{I}\right|  \right)  \otimes\left(  \left|  \alpha\right\rangle \left\langle
\alpha\right|  \right)  $. After coding (through the matrix $\widehat{C}$) it
reads $\widehat{\widetilde{\rho}}=\widehat{C}^{\dagger}\widehat{\rho}%
\widehat{C}$ ; at the end of the action of the errors it is transformed into
$\widehat{\widetilde{\rho}}_{e}=\widehat{U}_{inf}^{\dagger}\widehat
{C}^{\dagger}\widehat{\rho}\widehat{C}\widehat{U}_{inf}$ ; finally it takes
the form $\widehat{\rho}_{e}=\widehat{C}\widehat{U}_{inf}^{\dagger}\widehat
{C}^{\dagger}\widehat{\rho}\widehat{C}\widehat{U}_{inf}\widehat{C}^{\dagger}$
after decoding. In this setting, the projection onto $\mathcal{C}$ can be
simply achieved by measuring the ancilla in its initial state $\left|
\alpha\right\rangle $. As $T$ is very short, the state of the ancilla evolves
just a little within a Zeno interval, such that the probability of detecting
it in its initial state $\left|  \alpha\right\rangle $, and thus of projecting
the state of the compound system onto $\mathcal{C}$ is very close to $1$.
After projection, we trace out the ancilla to obtain the final reduced density
matrix $\widehat{\rho}_{I}^{\prime}=\left\langle \alpha\right|  \widehat
{C}\widehat{U}_{inf}^{\dagger}\widehat{C}^{\dagger}\widehat{\rho}\widehat
{C}\widehat{U}_{inf}\widehat{C}^{\dagger}\left|  \alpha\right\rangle $ for the
information system $\mathcal{I}$; in the same way, one can calculate the
initial reduced density matrix is $\widehat{\rho}_{I}=\left|
\psi_{I}\right\rangle \left\langle \psi_{I}\right|  .$ The variation
$\delta\widehat{\rho}_{I}=$ $\widehat{\rho}_{I}^{\prime}-\widehat{\rho}_{I}$
of the information-space density matrix during the whole process can be
expressed as the commutator
\[
\delta\widehat{\rho}_{I}=-i\left[  \sum_{m=1}^{M}\int f_{m}(\tau
)d\tau\left\langle \alpha\right|  \widehat{C}^{\dagger}\widehat{E}_{m}%
\widehat{C}\left|  \alpha\right\rangle ,\widehat{\rho}_{I}\right]  ,
\]
from which we infer that $\widehat{\rho}_{I}$ satisfies the equation
\begin{eqnarray}
i\frac{d\widehat{\rho}_{I}}{dt} & = & \left[  \widehat{h}_{e},\widehat{\rho
}_{I}\right]  , \nonumber \\
\widehat{h}_{e}& = & \sum_{m=1}^{M}f_{m}\left\langle
\alpha\right|  \widehat{C}^{\dagger}\widehat{E}_{m}\widehat{C}\left|
\alpha\right\rangle \nonumber
\end{eqnarray}
where $\widehat{h}_{e}$ is an effective Hamiltonian which is determined by the error-inducing
Hamiltonians transformed by the coding and decoding and projected onto the
initial state of the ancilla. From Eq.(\ref{matcod}) one can infer that 
$\widehat{h}_{e}=0$ and hence $\widehat{\rho}_{I}$\ remains constant in time:
as long as we repeat the coding-decoding-ancilla resetting sequence, the
information initially stored in $\mathcal{I}$\ is protected.

It is not always feasible to directly measure the ancilla
independently from the information system ; in other words, it is sometimes
impossible to perform a projection onto disentangled subspaces of $\mathcal{H}%
$\ of the form $\mathcal{H}_{I}\mathcal{\otimes}Span\left[  \left|
\alpha\right\rangle \right]  $\ : in some cases, as for the Rb atom (Sec.IV), one can only project onto entangled subspaces of
the total Hilbert space $\mathcal{H}$. In such a case the information
initially stored in the vector $\left|  \psi_{I}\right\rangle =\sum_{i=1}%
^{I}c_{i}\left|  \nu_{i}\right\rangle \in\mathcal{H}_{I}$ is transferred into
an entangled state of $\mathcal{I}$ and $\mathcal{A}$ of the form $\left|
\psi\right\rangle =$ $\sum_{i=1}^{I}c_{i}\left|  \gamma_{i}\right\rangle $
where the $I$ vectors $\left|  \gamma_{i}\right\rangle $ ($i=1,...,I$) which
form an orthonormal basis of the information-carrying subspace $\mathcal{C}$,
are not factorized as earlier but are in general entangled states. Nevertheless 
the same method as before can be used in that case to protect information, albeit in a 
different subspace $\mathcal{C}$.

To conclude this section, let us make a few remarks about our method. 
We first emphasize that our technique, though inspired by quantum error-correcting codes \cite{Chuang}, is very different from them : indeed, in those schemes, the information is encoded in such a way that it can be corrected from the action of a set of errors through a syndrome measurement, followed by a (conditioned) recovery operation, depending on the result of the measurement; on the other hand, in our technique, information is continuously protected by the frequent repetition of a three step cycle (coding-decoding-projective measurement), in which the projective measurement does not give any indication about which error occurred, but simply clears out the erroneous component of the state vector, which has been made orthogonal to the initial information-carrying subspace through coding and decoding.   
Let us now return to conditions
(\ref{bascod1}) and (\ref{bascod2})\ imposed on the codewords $\left\{
\left|  \widetilde{\gamma}_{i}\right\rangle ,i=1,...,I\right\}  $ and make two
points about them:

A. We can establish a useful relation between the dimension of the ancilla and
the number of correctable error Hamiltonians. The set of the $I$
codewords\ can be seen as a collection of $2I\times N=2I^{2}A$ real numbers on
which $2I^{2}+2MI^{2}=2I^{2}(1+M)$ constraints, directly derived from Eqs.(\ref{bascod1},\ref{bascod2}), are imposed. The number of free parameters must be larger
than the number of constraints, hence we necessarily have $2I^{2}A\geq2I^{2}(1+M)$, which satisfies
\begin{equation}
A-1\geq M. \label{Hamming}%
\end{equation}
This condition gives an upper-bound on the number of independent
error-inducing Hamiltonians that our method can correct simultaneously and is
called the ''Hamming bound''.

B. We may compare our correctability conditions (\ref{bascod2}) with the
more general conditions (see \cite{Chuang}\ p.436) of standard quantum
error-correction
\begin{eqnarray}
\forall\left(  \left|  \widetilde{\gamma}_{s}\right\rangle ,\left|
\widetilde{\gamma}_{t}\right\rangle \right)  \in\widetilde{\mathcal{C}}%
^{2},\text{ }\forall\left(  \widehat{\mathbf{E}}_{k},\widehat{\mathbf{E}}%
_{l}\right)  \in\left\{  \widehat{\mathbf{E}}_{j}\left(  \left\{  \widehat
{E}_{m}\right\}  \right)  \right\}  ,\nonumber \\ \left\langle \widetilde{\gamma
}_{t}\left|  \widehat{\mathbf{E}}_{k}^{\dagger}\widehat{\mathbf{E}}%
_{l}\right|  \widetilde{\gamma}_{s}\right\rangle =\alpha_{kl}\left\langle
\widetilde{\gamma}_{t}|\widetilde{\gamma}_{s}\right\rangle \label{correcgene}%
\end{eqnarray}
which ensure the existence of a code space that is completely protected against the
error-inducing Hamiltonians $\widehat{E}_{m}$. Here $\alpha_{kl}$ are complex
numbers, and the set $\left\{  \widehat{E}_{m}\right\}  $\ of Hermitian
operators $\widehat{E}_{m}$ generates a group $\mathcal{G}\left(  \left\{
\widehat{E}_{m}\right\}  \right)  $\ of all possible error-induced evolutions
(\ref{evolinf}). By $\left\{  \widehat{\mathbf{E}}_{j}\left(  \left\{
\widehat{E}_{m}\right\}  \right)  \right\}  $ we denote a complete basis set
of operators which spans the space of evolution operators $\widehat{U}$ and
allows one to represent any $\widehat{U}$ as a linear combination of the basis
operators $\widehat{\mathbf{E}}_{j}$. The variety of all linear combinations
of $\widehat{\mathbf{E}}_{j}$\ includes not only all $\widehat{E}_{m}$\ but
also many other operators given by commutators of all orders in $\widehat{E}_{m}$
entering the expansion of $\widehat{U}$ for
long times. The condition (\ref{correcgene}) is therefore much more
restrictive than Eq.(\ref{bascod2}). Moreover, even for just two generic
matrices $\widehat{E}_{m}$, the basis $\left\{  \widehat{\mathbf{E}}%
_{j}\right\}  $\ spans the entire Hilbert space $\mathcal{H}$, yielding
$\widetilde{\mathcal{C}}=\emptyset$. Only if the set $\left\{  \widehat
{E}_{m}\right\}  $\ belongs to an extraspecial algebra restricting the error
evolution operators $\widehat{U}$ to a subgroup $\mathcal{G}\left(  \left\{
\widehat{E}_{m}\right\}  \right)  \subset\mathcal{G}_{U}\left(  \mathcal{H}%
\right)  $ of the full unitary group in $\mathcal{H}$, a non-trivial code
space $\widetilde{\mathcal{C}}$\ may exist. The Zeno effect is the only way to
suppress loss of coherence if it is not the case.

\section{The code space and the coding matrix\label{III}}

It is sometimes possible to build the code space $\widetilde{\mathcal{C}}%
$\ explicitly from physical considerations: Appendix A gives an example of
a situation in which the code basis can be found directly. In general,
however, we need an algorithm to calculate the code basis $\left\{  \left|
\widetilde{\gamma}_{i}\right\rangle \right\}  _{i=1,...,I}$ or, equivalently,
the coding matrix $\widehat{C}$. We start this section by describing this
algorithm. Then, in a second part, we show that the non-holonomic control
technique \cite{NHC} can be employed to implement the coding matrix
physically. We also provide an algorithm which achieves the appropriate control.

Let us first make a remark which will be useful in what follows. Consider a vector $\left|  \mathsf{C}\right\rangle $ of some Hilbert
space and a matrix $\widehat{\mathsf{E}}$ on this space. From the vector
$\left|  \mathsf{C}\right\rangle $ we want to calculate a vector $\left|
\widetilde{\mathsf{C}}\right\rangle $\ such that $\left\langle \widetilde
{\mathsf{C}}\left|  \widehat{\mathsf{E}}\right|  \widetilde{\mathsf{C}%
}\right\rangle =0$. If $\left\langle \mathsf{C}\left|  \widehat{\mathsf{E}%
}\right|  \mathsf{C}\right\rangle =0$, then $\left|  \mathsf{C}\right\rangle
=\left|  \widetilde{\mathsf{C}}\right\rangle $ and the function
\[
f_{\widetilde{\mathsf{C}}}(\lambda)=\left\|  \left|  \widetilde{\mathsf{C}%
}\right\rangle +\lambda\widehat{\mathsf{E}}\left|  \widetilde{\mathsf{C}%
}\right\rangle \right\|  ^{2},%
\]
depending on the c-number $\lambda$, is minimal for $\lambda=0$: indeed
\begin{widetext}
\[
\left\|  \left|  \widetilde{\mathsf{C}}\right\rangle +\lambda\widehat
{\mathsf{E}}\left|  \widetilde{\mathsf{C}}\right\rangle \right\|  ^{2}=\left\langle \widetilde{\mathsf{C}}|\widetilde{\mathsf{C}}\right\rangle
+\lambda\left\langle \widetilde{\mathsf{C}}\left|  \widehat{\mathsf{E}
}\right|  \widetilde{\mathsf{C}}\right\rangle  +\lambda^{\ast}\left\langle
\widetilde{\mathsf{C}}\left|  \widehat{\mathsf{E}}^{\dagger}\right|
\widetilde{\mathsf{C}}\right\rangle +\left|  \lambda\right|  ^{2}\left\langle
\widetilde{\mathsf{C}}\left|  \widehat{\mathsf{E}}^{\dagger}\widehat
{\mathsf{E}}\right|  \widetilde{\mathsf{C}}\right\rangle  =1+\left|  \lambda\right|  ^{2}\left\langle \widetilde{\mathsf{C}}\left|
\widehat{\mathsf{E}}^{\dagger}\widehat{\mathsf{E}}\right|  \widetilde
{\mathsf{C}}\right\rangle ,
\]
\end{widetext}
and as $\left\langle \widetilde{\mathsf{C}}\left|  \widehat{\mathsf{E}%
}^{\dagger}\widehat{\mathsf{E}}\right|  \widetilde{\mathsf{C}}\right\rangle
\geq0$, $f_{\widetilde{\mathsf{C}}}(\lambda)$ is minimal for $\left|
\lambda\right|  =0$, that is $\lambda=0$. But if $\left\langle \mathsf{C}%
\left|  \widehat{\mathsf{E}}\right|  \mathsf{C}\right\rangle \neq0$, we can
apply the following iterative method: we minimize $f_{\mathsf{C}}(\lambda
)$\ with respect to $\lambda$, then we set $\left|  \mathsf{C}^{\prime
}\right\rangle =\left|  \mathsf{C}\right\rangle +\frac{\lambda}{2}%
\widehat{\mathsf{E}}\left|  \mathsf{C}\right\rangle $ and take $\frac{\left|
\mathsf{C}^{\prime}\right\rangle }{\sqrt{\left\langle \mathsf{C}^{\prime
}|\mathsf{C}^{\prime}\right\rangle }}$ as our new $\left|  \mathsf{C}%
\right\rangle $ ; we repeat this sequence as long as needed: $\left|
\mathsf{C}\right\rangle $ finally tends to $\left|  \widetilde{\mathsf{C}%
}\right\rangle $, such that $\left\langle \widetilde{\mathsf{C}}\left|
\widehat{\mathsf{E}}\right|  \widetilde{\mathsf{C}}\right\rangle =0$.

Let us now return to our problem and show how the previous remark can help us. What we
want is to find $I$ vectors $\left|  \widetilde{\gamma}_{i}\right\rangle $\ which
meet the conditions (\ref{bascod1}) and (\ref{bascod2}) ; equivalently, we can say that we look for an orthonormal basis in which all the matrices $\widehat{\mathsf{E}}_{k}$ have their $I \times I$ upper left blocks equal to zero. To solve this problem, one can first be tempted to use standard techniques of linear algebra, in particular matrix diagonalization : however, it appears that these methods do not work, except in the trivial case when all the matrices $\widehat{\mathsf{E}}_{k}$ have a common kernel, which is much more than what conditions (\ref{bascod1}) and (\ref{bascod2}) require. So we propose to transform our initial problem in such a way that it can be dealt with by the iterative algorithm presented in the previous paragraph. Let us combine
the $I$ vectors $\left|  \widetilde{\gamma}_{i}\right\rangle $ into a $\left(
N\times I\right)  $ ''supervector''
\[
\left|  \widetilde{\mathsf{C}}\right\rangle =\left(
\begin{array}
[c]{c}%
\left|  \widetilde{\gamma}_{1}\right\rangle \\
\vdots\\
\left|  \widetilde{\gamma}_{I}\right\rangle
\end{array}
\right)  .
\]
Then let us build $E=\left(  \frac{I(I-1)}{2}+M\frac{I(I+1)}{2}\right)  $
different $\left(  N\times I\right)  \times\left(  N\times I\right)
$-dimensional super-matrices $\widehat{\mathsf{E}}_{k}$ in the following way:
we consider them as made of $I^{2}$ blocks of dimension $N\times N$ and we
successively fill each of these blocks with the different Hamiltonians
$\widehat{E}_{m}$ or the identity matrix $\widehat{I}$ or $0$. To be more
explicit, the first $\frac{I(I-1)}{2}$ matrices are built by simply placing
the $N\times N$ identity matrix in each of the $\frac{I(I-1)}{2}$ blocks
situated above the diagonal. In the last $\frac{MI(I+1)}{2}$ ones, the $M$
operators $\widehat{E}_{m}$ are successively placed in each of the
$\frac{I(I+1)}{2}$ blocks on and above the diagonal. One can thus reformulate
the conditions (\ref{bascod1}) as follows: for $1\leq k\leq\frac{I(I-1)}{2}$
\[\left\langle \widetilde
{\mathsf{C}}\left|  \widehat{\mathsf{E}}_{k}\right|  \widetilde{\mathsf{C}%
}\right\rangle =0.
\]
Note that this form does not take the normalization condition into account,
which will be imposed differently. 
Similarly, the conditions (\ref{bascod2}) are translated into the following form: for $\frac{I(I-1)}{2}+1\leq k\leq\frac{I(I-1)}{2}+\frac{MI(I+1)}%
{2}$,
\[\left\langle \widetilde{\mathsf{C}}\left|  \widehat
{\mathsf{E}}_{k}\right|  \widetilde{\mathsf{C}}\right\rangle =0.
\]
Our initial multivectorial problem given by Eq.(\ref{bascod1},\ref{bascod2}) has thus been transformed\ into a simpler one which can
be handled by the same kind of iterative algorithm as in our preliminary
remark: we just have to find a $\left(  N\times I\right)  $-dimensional
supervector $\left|  \widetilde{\mathsf{C}}\right\rangle $ such that for $1\leq k\leq\frac{I(I-1)}{2}+\frac{MI(I+1)}{2}$,
\[\left\langle \widetilde{\mathsf{C}}\left|  \widehat{\mathsf{E}}_{k}\right|
\widetilde{\mathsf{C}}\right\rangle =0.
\]

Let us now review our iterative algorithm in more detail. First we randomly
pick a supervector $\left|  \mathsf{C}_{0}\right\rangle $ which will be the
starting point of the first step: we normalize this vector by imposing to
each of its $I$ components to have norm = $\frac{1}{I}$. If one of the
components of $\left|  \mathsf{C}_{0}\right\rangle $ is non normalizable, that
is equals zero, we pick up a new random supervector $\left|  \mathsf{C}%
_{0}\right\rangle $ as a starting point.

Then, as in our preliminary remark, we minimize the function
\[
F_{\mathsf{C}_{0}}\left(  \lambda_{1}^{(0)},\lambda_{2}^{(0)},...,\lambda
_{E}^{(0)}\right)  =\sum_{k=1}^{E}\left\|  \left|  \mathsf{C}_{0}\right\rangle
+\lambda_{k}^{(0)}\widehat{\mathsf{E}}_{k}\left|  \mathsf{C}_{0}\right\rangle
\right\|  ^{2}%
\]
with respect to the $E$ c-numbers $\lambda_{k}^{(0)}$: actually, we separate
the real and imaginary parts of $\lambda_{k}^{(0)}=\alpha_{k}^{(0)}+i\beta
_{k}^{(0)}$\ and calculate the appropriate $\alpha_{k}^{(0)}$'s and $\beta
_{k}^{(0)}$'s\ by solving the set of $2E$ equations
\begin{eqnarray}
\frac{\partial F_{\mathsf{C}_{0}}}{\partial\alpha_{k}^{(0)}}=0 \nonumber \\
\frac{\partial F_{\mathsf{C}_{0}}}{\partial\beta_{k}^{(0)}}=0, \nonumber
\end{eqnarray}
which can be translated into the linear system
\[
\widehat{K}\left(  \left|  \mathsf{C}_{0}\right\rangle \right)
.\overrightarrow{\Lambda}^{(0)}=\overrightarrow{D}\left(  \left|
\mathsf{C}_{0}\right\rangle \right)
\]
where $\widehat{K}\left(  \left|  \mathsf{C}_{0}\right\rangle \right)  $ is a
$2E\times2E$-dimensional real matrix defined by
\begin{widetext}
\[
\widehat{K}_{ij}\left(  \left|  \mathsf{C}_{0}\right\rangle \right)
=\left\{
\begin{array}
[c]{c}%
\mbox{Re}\left(  \left\langle \mathsf{C}_{0}\left|  \widehat
{\mathsf{E}}_{i}^{\dagger}\widehat{\mathsf{E}}_{j}\right|  \mathsf{C}%
_{0}\right\rangle \right)  \text{ \ \ for }1\leq i\leq E\text{ and }1\leq
j\leq E\\
-\mbox{Im}\left(  \left\langle \mathsf{C}_{0}\left|  \widehat
{\mathsf{E}}_{i}^{\dagger}\widehat{\mathsf{E}}_{j-E}\right|  \mathsf{C}%
_{0}\right\rangle \right)  \text{ \ \ for }1\leq i\leq E\text{ and }1+E\leq
j\leq2E\\
\mbox{Im}\left(  \left\langle \mathsf{C}_{0}\left|  \widehat
{\mathsf{E}}_{i-E}^{\dagger}\widehat{\mathsf{E}}_{j}\right|  \mathsf{C}%
_{0}\right\rangle \right)  \text{ \ \ for }1+E\leq i\leq2E\text{ and }1\leq
j\leq E\\
\mbox{Re}\left(  \left\langle \mathsf{C}_{0}\left|  \widehat
{\mathsf{E}}_{i-E}^{\dagger}\widehat{\mathsf{E}}_{j-E}\right|  \mathsf{C}%
_{0}\right\rangle \right)  \text{ \ \ for }1+E\leq i\leq2E\text{ and }1+E\leq
j\leq2E
\end{array}
\right.  ,
\]
\end{widetext}
$\overrightarrow{D}\left(  \left|  \mathsf{C}_{0}\right\rangle \right)  $ is a
$2E$-dimensional real vector defined by
\[
\overrightarrow{D}\left(  \left|  \mathsf{C}_{0}\right\rangle \right)
=\left\{
\begin{array}
[c]{c}%
-\mbox{Re}\left(  \left\langle \mathsf{C}_{0}\left|  \widehat
{\mathsf{E}}_{i}\right|  \mathsf{C}_{0}\right\rangle \right)  \text{ \ \ for
}1\leq i\leq E\\
\mbox{Im}\left(  \left\langle \mathsf{C}_{0}\left|  \widehat
{\mathsf{E}}_{i-E}\right|  \mathsf{C}_{0}\right\rangle \right)  \text{ \ \ for
}E+1\leq i\leq2E
\end{array}
\right.
\]
and $\overrightarrow{\Lambda}^{(0)}$ is a $2E$-dimensional real vector containing the
parameters $\alpha_{k}^{(0)}$'s and $\beta_{k}^{(0)}$'s
\[
\overrightarrow{\Lambda}^{(0)}=\left(
\begin{array}
[c]{c}%
\alpha_{1}^{(0)}\\
\vdots\\
\alpha_{E}^{(0)}\\
\beta_{1}^{(0)}\\
\vdots\\
\beta_{E}^{(0)}%
\end{array}
\right)  .
\]
Once the c-numbers $\lambda_{k}^{(0)}=\left(  \alpha_{k}^{(0)}+i\beta_{k}^{(0)}\right)
$'s have been found, we calculate $\left|  \Delta\mathsf{C}_{0}\right\rangle
=\sum_{k}\lambda_{k}^{(0)}\widehat{\mathsf{E}}_{k}\left|  \mathsf{C}%
_{0}\right\rangle $ and $\left|  \mathsf{C}_{0}^{\prime}\right\rangle =\left|
\mathsf{C}_{0}\right\rangle +\frac{1}{2}\left|  \Delta\mathsf{C}%
_{0}\right\rangle $. We normalize $\left|  \mathsf{C}_{0}^{\prime
}\right\rangle $ by requiring each of its $I$ components to have the norm =
$\frac{1}{I}$, and take the result of this operation as our new starting point
$\left|  \mathsf{C}_{1}\right\rangle $. If one of the components of $\left|
\mathsf{C}_{0}^{\prime}\right\rangle $ is non normalizable, that is equals
zero, we pick up a new random supervector $\left|  \mathsf{C}_{0}\right\rangle
$ as a starting point.

We repeat this sequence of operations as long as needed. Thus, at the $m$th
step$,$ we minimize the function
\begin{eqnarray*}
F_{\mathsf{C}_{m-1}}\left(  \lambda_{1}^{\left(  m-1\right)  },\lambda
_{2}^{\left(  m-1\right)  },...,\lambda_{E}^{\left(  m-1\right)  }\right)
\\ =\sum_{k=1}^{E}\left\|  \left|  \mathsf{C}_{m-1}\right\rangle +\lambda
_{k}^{\left(  m-1\right)  }\widehat{\mathsf{E}}_{k}\left|  \mathsf{C}%
_{m-1}\right\rangle \right\|  ^{2}%
\end{eqnarray*}
by solving the real linear system
\[
\widehat{K}\left(  \left|  \mathsf{C}_{m-1}\right\rangle \right)
.\overrightarrow{\Lambda}^{\left(  m-1\right)  }=\overrightarrow{D}\left(
\left|  \mathsf{C}_{m-1}\right\rangle \right)  .
\]
This yields the $\lambda_{k}^{\left(  m-1\right)  }$'s and $\left|
\Delta\mathsf{C}_{m-1}\right\rangle $ from which we calculate $\left|
\mathsf{C}_{m-1}^{\prime}\right\rangle =\left|  \mathsf{C}_{m-1}\right\rangle
+\frac{1}{2}\left|  \Delta\mathsf{C}_{m-1}\right\rangle $. If possible,
we normalize $\left|  \mathsf{C}_{m-1}^{\prime}\right\rangle $\ and take the
resulting vector as the starting point $\left|  \mathsf{C}_{m}\right\rangle
$\ of the $(m+1)$th step ; otherwise, we pick up a new vector $\left|
\mathsf{C}_{0}\right\rangle $\ as a starting point. Finally $\left|  \mathsf{C}_{m}\right\rangle $ tends to $\left|
\widetilde{\mathsf{C}}\right\rangle $ such that $\forall k\in\left[
1,\frac{I(I-1)}{2}\right]  ,\left\langle \widetilde{\mathsf{C}}\left|
\widehat{\mathsf{E}}_{k}\right|  \widetilde{\mathsf{C}}\right\rangle =0.$

This algorithm was numerically implemented and allowed us to exhibit new codes
: we protected $2$ qubits among $7$ against the action of $31$ errors ($21$
individual errors + $10$ collective errors) and $4$ qubits among $9$ against
the action of $27$ individual errors \cite{CPZE}.

\bigskip

\bigskip

The coding matrix $\widehat{C}$ which allows us to transfer the information from
the space $\mathcal{C}$ to the code space $\widetilde{\mathcal{C}}$, is a
rather complex unitary operator on the Hilbert space of the compound
system $\mathcal{S=I}\otimes\mathcal{A}$. We have just shown how to calculate 
the codewords, which actually form the first $I$ columns
of $\widehat{C}$, but one can wonder how to implement it physically. The
question of the physical feasibility of the coding matrix $\widehat{C}$ can be
solved by the non-holonomic control technique.

The non-holonomic control technique has been suggested by our team as a means
of controlling the evolution of quantum systems \cite{NHC}. Basically, it
consists in alternately applying two ''well-chosen''
perturbations $\widehat{V}_{a}$ and $\widehat{V}_{b}$ to the system
$\mathcal{S}$ we want to control during pulses with timings $t_{i}$. The total
Hamiltonian $\widehat{H}=\widehat{H}_{0}+\widehat{V}$ thus has a pulsed shape
and alternately takes the two values $\widehat{H}_{a}\equiv\widehat{H}%
_{0}+\widehat{V}_{a}$ (during odd-numbered pulses) and $\widehat{H}%
_{b}\equiv\widehat{H}_{0}+\widehat{V}_{b}$ (even-numbered pulses).
The timings $t_{i}$ play the role of free parameters one has to adjust in order to
perform the desired control operation. To be more explicit, the perturbations $\widehat{V}_{a}$ and
$\widehat{V}_{b}$ must be chosen so that the commutators of all orders of
$\widehat{H}_{a}\equiv\widehat{H}_{0}+\widehat{V}_{a}$ and $\widehat{H}%
_{b}\equiv\widehat{H}_{0}+\widehat{V}_{b}$ span the whole space of Hermitian
matrices acting on the system we want to control: this is called the
\emph{bracket generation condition} (BGC). From the Campbell-Baker-Hausdorf
formula, it follows that this is a necessary condition of
controllability. It also proves to be sufficient in all the practical
cases we dealt with. For that reason, we consider that we have ''good
controllability conditions'' as soon as BGC is checked.
The number $n_{C}$ of control timings depends on the problem to be solved. 
For instance, if we want to impose the arbitrary
evolution $\widehat{U}_{arb}$ on an $N$-dimensional system, we need at least
$n_{C}=N^{2}$ timings $t_{i}$, since $N^{2}$ is the total number of free real
parameters characterizing a $N\times N$ unitary matrix. We dealt with this
problem of complete control in previous papers \cite{NHC}, and developed a
general algorithm to find the appropriate timings $t_{i}$ which realize
\begin{eqnarray*}
\widehat{U}\left(  t_{1},t_{2},...,t_{N^{2}}\right)  =\exp\left(
-i\widehat{H}_{a}t_{N^{2}}\right)  \exp\left(  -i\widehat{H}_{b}%
t_{N^{2}-1}\right) \\  \ldots\exp\left(  -i\widehat{H}_{b}t_{1}\right)
=\widehat{U}_{arb}.
\end{eqnarray*}

\bigskip 
We can directly apply this result to our coding problem in the following 
way: first, we find the codewords $\left\{  \left|
\widetilde{\gamma}_{i}\right\rangle ,i=1,...,I\right\}  $ by the iterative
algorithm we have presented in the first part of this section, then we complete
the set of $I$ vectors $\left\{  \left|  \widetilde{\gamma}_{i}\right\rangle
,i=1,...,I\right\}  $ with $\left(  N-I\right)  $ vectors $\left\{  \left|
\widetilde{\gamma}_{j}\right\rangle ,j=I+1,...,N\right\}  $ to form an
orthonormal basis of $\mathcal{H}$, we buil\-d the coding matrix by taking the
vectors $\left\{  \left|  \widetilde{\gamma}_{i}\right\rangle
,i=1,...,N\right\}  $\ as columns of $\widehat{C}$, and finally we calculate
the $n_{C}=N^{2}$ appropriate timings $\left\{  t_{i}\right\}  $ such that%

\begin{eqnarray*}
\widehat{U}\left(  t_{1},t_{2},...,t_{N^{2}}\right)  =\exp\left(
-i\widehat{H}_{a}t_{N^{2}}\right)  \exp\left(  -i\widehat{H}_{b}%
t_{N^{2}-1}\right) \\ \ldots\exp\left(  -i\widehat{H}_{b}t_{1}\right)
=\widehat{C}%
\end{eqnarray*}
through the complete control algorithm presented in \cite{NHC}. Note that we assume $\widehat
{H}_{0}=0$ (Sec.II), hence $\widehat
{H}_{a}=\widehat
{V}_{a}$ and $\widehat
{H}_{b}=\widehat
{V}_{b}$

Actually, this procedure provides a lot of useless work: indeed,
most of the information contained in the coding matrix is irrelevant and the
$N^{2}$\ real parameters of $\widehat{C}$\ do not all have to be controlled
exactly: the number $n_{C}$\ of necessary control parameters
$\left\{  t_{i}\right\}  $\ is much less than $N^{2}$. Let us examine this
point in more detail.

The coding matrix is characterized by the relations (\ref{matcodbis}). The
problem of control thus reduces to finding $n_{C}$\ timings $t_{i}$, which we
will formally gather in a time-vector $\overrightarrow{t}=\left(
\begin{array}
[c]{c}%
t_{1}\\
\vdots\\
t_{n_{C}}%
\end{array}
\right)  $, such that the non-holonomic evolution matrix
\begin{eqnarray*}
\widehat{U}\left(  \overrightarrow{t}\right)  =\exp\left(  -i\widehat{H}%
_{a}t_{n_{C}}\right) \exp\left(  -i\widehat{H}_{b}t_{n_{C}-1}\right) \\
\ldots\exp\left(  -i\widehat{H}_{a}t_{1}\right)
\end{eqnarray*}
meets conditions (\ref{matcodbis}): for any pair $\left(
\left|  \gamma_{s}\right\rangle ,\left|  \gamma_{t}\right\rangle \right)
_{1\leq s,t\leq I}$ of basis vectors of $\mathcal{C}$ and any operator $\widehat{E}_{m}%
\in\left\{  \widehat{E}_{m}\right\}  _{m=1,...,M}$%
\begin{equation}
\left\langle \gamma_{t}\left|  \widehat{U}^{\dagger}\left(  \overrightarrow
{t}\right)  \widehat{E}_{m}\widehat{U}\left(  \overrightarrow{t}\right)
\right|  \gamma_{s}\right\rangle =0. \label{matcodter}%
\end{equation}
The number $n_{C}$ of control parameters must exceed the number of
independent constraints which is clearly $\sim MI^{2}$, that is $n_{C}\gtrsim
MI^{2}$. The number of really necessary control parameters appears to be much
smaller than $N^{2}$. We have to design a new algorithm which achieves a
partial and less expensive control of the evolution operator of the system.

The algorithm we shall use to calculate the appropriate control timings
$t_{i}$ mixes the iterative algorithm presented at the beginning of this
section and the non-holonomic control technique. If we introduce the $\left(
N\times I\right)  \times\left(  N\times I\right)  $-dimensional block-diagonal
matrix
\[
\widehat{\mathsf{U}}\left(  \overrightarrow{t}\right)  =\left(
\begin{array}
[c]{cccc}%
\widehat{U}\left(  \overrightarrow{t}\right)  & 0 & \cdots & 0\\
0 & \widehat{U}\left(  \overrightarrow{t}\right)  & \cdots & 0\\
\vdots & \vdots & \vdots & \vdots\\
0 & 0 & \cdots & \widehat{U}\left(  \overrightarrow{t}\right)
\end{array}
\right)
\]
and the $\left(N\times I\right)$-dimensional supervector
\[
\left|  \mathsf{C}\right\rangle =\left(
\begin{array}
[c]{c}%
\left|  \gamma_{1}\right\rangle \\
\vdots\\
\left|  \gamma_{I}\right\rangle
\end{array}
\right)
\]
composed of the coordinates of the $I$ basis vectors of $\mathcal{C}$, we can
set the problem of control Eq.(\ref{matcodter}) in the following equivalent
form: we look for a time-vector $\overrightarrow{t}$ such that
\begin{equation}
\forall k,\text{ \ }\left\langle \mathsf{C}\left|  \widehat{\mathsf{U}%
}^{\dagger}\left(  \overrightarrow{t}\right)  \widehat{\mathsf{E}}_{k}%
\widehat{\mathsf{U}}\left(  \overrightarrow{t}\right)  \right|  \mathsf{C}%
\right\rangle =0 \label{matcodquattro}%
\end{equation}
where the matrices $\left\{  \widehat{\mathsf{E}}_{k}\right\}  _{k=1,...,E}%
$\ denote $E$ different matrices of dimension $\left(  N\times I\right)  \times\left(  N\times
I\right)  $ which have been introduced in the beginning
of this section. In other words, we look for the time-vector $\overrightarrow
{t}$\ which sets to zero the test function $G$($\overrightarrow{t})\equiv\sum
_{k=1}^{E}\left|  \left\langle \mathsf{C}\left|  \widehat{\mathsf{U}}%
^{\dagger}\left(  \overrightarrow{t}\right)  \widehat{\mathsf{E}}_{k}%
\widehat{\mathsf{U}}\left(  \overrightarrow{t}\right)  \right|  \mathsf{C}%
\right\rangle \right|  ^{2}$. The idea of our algorithm is to take the super
vector $\left|  \mathsf{C}_{0}\right\rangle =\widehat{\mathsf{U}}\left(
\overrightarrow{t}_{0}\right)  \left|  \mathsf{C}\right\rangle $, where
$\overrightarrow{t}_{0}$ is a random time-vector, as the starting point for an
elementary step of the iterative algorithm and look for the small time
increment $\overrightarrow{dt}_{0}$ such that $\widehat{\mathsf{U}}\left(
\overrightarrow{t}_{0}+\overrightarrow{dt}_{0}\right)  \left|  \mathsf{C}%
\right\rangle $\ follows the direction provided by the result $\left|
\mathsf{C}_{0}\right\rangle +\left|  \Delta\mathsf{C}_{0}\right\rangle $ of
the iterative algorithm. The repetition of this sequence finally yields
$\overrightarrow{t}=\overrightarrow{t}_{0}+\overrightarrow{dt}_{0}%
+\overrightarrow{dt}_{1}+...$ which meets Eq.(\ref{matcodquattro}).

Let us now describe the algorithm in more detail. First, we randomly pick a
set of timings $t_{0,i}$ in a ''realistic range'', dictated by the system under
consideration: in particular, control-pulse timings have to be much shorter
than the typical lifetime of the system and be much longer than the typical response delay required
by the experiment. Then we minimize the function
\[
F_{\mathsf{C}_{0}}\left(  \lambda_{1}^{\left(  0\right)  },\lambda
_{2}^{\left(  0\right)  },...,\lambda_{E}^{\left(  0\right)  }\right)
=\sum_{k=1}^{E}\left\|  \left|  \mathsf{C}_{0}\right\rangle +\lambda
_{k}^{\left(  0\right)  }\widehat{\mathsf{E}}_{k}\left|  \mathsf{C}%
_{0}\right\rangle \right\|  ^{2}%
\]
as we did in the algorithm presented at the beginning of this section: we
obtain the $\lambda_{k}^{\left(  0\right)  }$'s and $\left|  \Delta
\mathsf{C}_{0}\right\rangle =\sum_{k}\lambda_{k}\widehat{\mathsf{E}}%
_{k}\left|  \mathsf{C}_{0}\right\rangle $. At that point, we look for the
small increment $\overrightarrow{dt}_{0}$ of the time-vector $\overrightarrow
{t}_{0}$ such that
\begin{widetext}
\begin{eqnarray}
\forall k,\left\langle \mathsf{C}\left|  \left(  \frac{\partial
\widehat{\mathsf{U}}^{\dagger}}{\partial\overrightarrow{t}}\left(
\overrightarrow{t}_{0}\right)  .\overrightarrow{dt}_{0}\right)  \widehat
{\mathsf{E}}_{k}\widehat{\mathsf{U}}\left(  \overrightarrow{t}_{0}\right)
+\widehat{\mathsf{U}}^{\dagger}\left(  \overrightarrow{t}_{0}\right)
\widehat{\mathsf{E}}_{k}\left(  \frac{\partial\widehat{\mathsf{U}}}%
{\partial\overrightarrow{t}}\left(  \overrightarrow{t}_{0}\right)
.\overrightarrow{dt}_{0}\right)  \right|  \mathsf{C}\right\rangle \nonumber \\
=\frac{\left\langle \mathsf{C}_{0}+\frac{1}{2}\Delta\mathsf{C}_{0}\left|
\widehat{\mathsf{E}}_{k}\right|  \mathsf{C}_{0}+\frac{1}{2}\Delta
\mathsf{C}_{0}\right\rangle -\left\langle \mathsf{C}_{0}\left|  \widehat
{\mathsf{E}}_{k}\right|  \mathsf{C}_{0}\right\rangle }{\left\langle
\mathsf{C}_{0}+\frac{1}{2}\Delta\mathsf{C}_{0}|\mathsf{C}_{0}+\frac{1}%
{2}\Delta\mathsf{C}_{0}\right\rangle }. \label{equation}%
\end{eqnarray}
\end{widetext}

It should be noticed that we do not consider the error super-matrices
$\widehat{\mathsf{E}}_{k}$ corresponding to orthonormality conditions: in other words, we just take matrices $\left\{  \widehat{\mathsf{E}}_{k}\right\}
_{k\in\left[  \frac{I(I-1)}{2}+1,\frac{I(I-1)}{2}+\frac{MI(I+1)}{2}\right]  }$
into account. Thus we deal with $\frac{MI(I+1)}{2}$ complex equations. This
set of equations can be reduced to the real linear system
\begin{equation}
\widehat{S}\left(  \overrightarrow{t}_{0}\right)  \cdot\overrightarrow{dt}%
_{0}=\overrightarrow{W}\left(  \left|  \Delta\mathsf{C}_{0}\right\rangle
\right)  \label{systeme}%
\end{equation}
where $\widehat{S}\left(  \overrightarrow{t}_{0}\right)  $ and
$\overrightarrow{W}\left(  \left|  \Delta\mathsf{C}_{0}\right\rangle \right)
$ are respectively an $MI%
{{}^2}%
\times n_{C}$ real matrix and a $MI%
{{}^2}%
$-dimensional real vector. We obtained Eq.(\ref{systeme}) by splitting the set of $\frac{MI(I+1)}{2}
$ complex equations (\ref{equation}) into two sets of $\frac{MI(I+1)}{2}$
real equations, and rejecting those which are trivial ($0=0$) or redundant. Even though this procedure is straightforward, the explicit expressions of the different elements of $\widehat{S}$ and $\overrightarrow{W}$ involve many indices and are so unpleasant that we prefer not to reproduce them here.

The linear system we have just found is a priori rectangular $(MI%
{{}^2}%
\times n_{C})$, but actually we have not fixed the number $n_{C}$ yet.
Previously, we stated that $n_{C}\geq MI%
{{}^2}%
$: we could be tempted to set $n_{C}=MI%
{{}^2}%
$ so as to obtain a square system, easily solvable by standard techniques of
linear algebra. Yet we will proceed in a slightly different way. We
set $n_{C}>MI%
{{}^2}%
$, say $n_{C}=MI%
{{}^2}%
+\delta n$ where $\delta n$ is an integer of order $1$. Then we randomly pick
$MI%
{{}^2}%
$ timings $t_{i}$ among the $n_{C}$ which will be considered as free
parameters, whereas the other $\delta n$ ones will be regarded as frozen. In
other words, we randomly choose a permutation $\sigma_{0}\in\mathcal{S}%
_{n_{C}}$ (symmetric group of order $n_{C}$) and take the timings $\left\{
t_{i}^{\prime}=t_{\sigma_{0}(i)}\right\}  _{i=1,...,MI^{2}}$ as free
parameters whereas the timings $\left\{  t_{i}^{\prime}=t_{\sigma_{0}%
(i)}\right\}  _{i=1+MI^{2},n_{C}}$\ are frozen. This leads to new versions of
Eqs(\ref{equation},\ref{systeme}):
\begin{widetext}
\begin{eqnarray}
\forall k,\left\langle \mathsf{C}\left|  \left(  \frac{\partial
\widehat{\mathsf{U}}^{\dagger}}{\partial\overrightarrow{t^{\prime}}}\left(
\overrightarrow{t}_{0}\right)  \cdot\overrightarrow{dt^{\prime}}_{0}\right)
\widehat{\mathsf{E}}_{k}\widehat{\mathsf{U}}\left(  \overrightarrow{t}%
_{0}\right)  +\widehat{\mathsf{U}}^{\dagger}\left(  \overrightarrow{t}%
_{0}\right)  \widehat{\mathsf{E}}_{k}\left(  \frac{\partial\widehat
{\mathsf{U}}}{\partial\overrightarrow{t^{\prime}}}\left(  \overrightarrow
{t}_{0}\right)  \cdot\overrightarrow{dt^{\prime}}_{0}\right)  \right|
\mathsf{C}\right\rangle \nonumber \\ 
= \frac{\left\langle \mathsf{C}_{0}+\frac{1}{2}\Delta\mathsf{C}_{0}\left|
\widehat{\mathsf{E}}_{k}\right|  \mathsf{C}_{0}+\frac{1}{2}\Delta
\mathsf{C}_{0}\right\rangle -\left\langle \mathsf{C}_{0}\left|  \widehat
{\mathsf{E}}_{k}\right|  \mathsf{C}_{0}\right\rangle }{\left\langle
\mathsf{C}_{0}+\frac{1}{2}\Delta\mathsf{C}_{0}|\mathsf{C}_{0}+\frac{1}%
{2}\Delta\mathsf{C}_{0}\right\rangle }, \label{equation2}
\end{eqnarray}
\end{widetext}
\begin{equation}
\widehat{S}\left(  \overrightarrow{t}_{0}\right)  \cdot\overrightarrow
{dt^{\prime}}_{0} =\overrightarrow{W}\left(  \left|  \Delta\mathsf{C}%
_{0}\right\rangle \right)  . \label{systeme2}%
\end{equation}

Eq.(\ref{systeme2}) is now clearly a square system. Solving Equation (\ref{systeme2}%
) yields the $MI%
{{}^2}%
$-dimensional increment $\overrightarrow{dt^{\prime}}_{0}$ which we complete
with $\delta n$ zeros into a $n_{C}$-dimensional vector ; by reordering
timings we obtain the total time-vector increment $\overrightarrow{dt}_{0}$.
Thus we have for $i\in\left[  1,MI^{2}\right]  $, $dt_{0,\sigma_{0}(i)}\neq0$
(free parameters), whereas for $i\in\left[  1+MI^{2},n_{C}\right]  $,
$dt_{0,\sigma_{0}(i)}=0$ (frozen timings). Then we set $\overrightarrow{t}%
_{1}^{\alpha}$ $=\overrightarrow{t}_{0}+$ $\alpha$ $\overrightarrow{dt}_{0}$
where $\alpha$ is a convergence coefficient and calculate the test function
$G\left(  \overrightarrow{t}\right)  =\sum_{k}\left|  \left\langle
\mathsf{C}\left|  \widehat{\mathsf{U}}^{\dagger}\left(  \overrightarrow
{t}\right)  \widehat{\mathsf{E}}_{k}\widehat{\mathsf{U}}\left(
\overrightarrow{t}\right)  \right|  \mathsf{C}\right\rangle \right|  ^{2}$ in
$\overrightarrow{t}=\overrightarrow{t}_{1}^{\alpha}$ for different values of
$\alpha\in\left[  0,1\right]  $. If we find an $\alpha_{1}$ such that
$G\left(  \overrightarrow{t}_{1}^{\alpha}\right)  <G\left(  \overrightarrow
{t}_{0}\right)  $, we take $\overrightarrow{t}_{1}\equiv\overrightarrow{t}%
_{1}^{\alpha_{1}}$ as our new time-vector, and keep the same free-varying
timings: in other words, the permutation $\sigma_{1}$ governing the timings
that play the role of control parameters in the second step of the algorithm
remains the same, that is $\sigma_{1}=\sigma_{0}$. If we
cannot find an appropriate $\alpha_{1}$, this means we are situated in a local
minimum of $G $ ; then we set $\overrightarrow{t}_{1}\equiv\overrightarrow
{t}_{0}$ and pick a new set of free varying parameters by simply choosing a
new permutation $\sigma_{1}\neq\sigma_{0}$ randomly. This rotation procedure
among control parameters allows us to avoid possible local minima of the test
function $G$ we want to cancel.

We repeat this sequence of operations as long as needed. At the
$m$th\ step, we take the supervector $\left|  \mathsf{C}_{m-1}\right\rangle
=\widehat{\mathsf{U}}\left(  \overrightarrow{t}_{m-1}\right)  \left|
\mathsf{C}\right\rangle $ as the starting point of an elementary step of the
iterative algorithm. We calculate $\left|  \Delta\mathsf{C}_{m-1}%
\right\rangle =\sum\lambda_{k}^{\left(  m-1\right)  }\widehat{\mathsf{E}}%
_{k}\left|  \mathsf{C}_{m-1}\right\rangle $ and find $MI%
{{}^2}%
$-dimensional variations vector $\overrightarrow{dt^{\prime}}_{m-1}$ of the
$MI%
{{}^2}%
$ free parameters (characterized by permutation $\sigma_{m-1}$) such that

\begin{widetext}
\begin{eqnarray*}
\forall k,\left\langle \mathsf{C}\left|  \left(  \frac{\partial
\widehat{\mathsf{U}}^{\dagger}}{\partial\overrightarrow{t^{\prime}}}\left(
\overrightarrow{t}_{m}\right)  \cdot\overrightarrow{dt^{\prime}}_{m}\right)
\widehat{\mathsf{E}}_{k}\widehat{\mathsf{U}}\left(  \overrightarrow{t}%
_{m}\right)  +\widehat{\mathsf{U}}^{\dagger}\left(  \overrightarrow{t}%
_{m}\right)  \widehat{\mathsf{E}}_{k}\left(  \frac{\partial\widehat
{\mathsf{U}}}{\partial\overrightarrow{t^{\prime}}}\left(  \overrightarrow
{t}_{m}\right)  \cdot\overrightarrow{dt^{\prime}}_{m}\right)  \right|
\mathsf{C}\right\rangle \\
=\frac{\left\langle \mathsf{C}_{m-1}+\frac{1}{2}\Delta\mathsf{C}%
_{m-1}\left|  \widehat{\mathsf{E}}_{k}\right|  \mathsf{C}_{m-1}+\frac{1}%
{2}\Delta\mathsf{C}_{m-1}\right\rangle -\left\langle \mathsf{C}_{m-1}\left|
\widehat{\mathsf{E}}_{k}\right|  \mathsf{C}_{m-1}\right\rangle }{\left\langle
\mathsf{C}_{m-1}+\frac{1}{2}\Delta\mathsf{C}_{m-1}|\mathsf{C}_{m-1}+\frac
{1}{2}\Delta\mathsf{C}_{m-1}\right\rangle }
\end{eqnarray*}
\end{widetext}
by solving the associated square linear system
\[
\widehat{S}\left(  \overrightarrow{t}_{m-1}\right)  \cdot\overrightarrow
{dt^{\prime}}_{m-1}=\overrightarrow{W}\left(  \left|  \Delta\mathsf{C}%
_{m-1}\right\rangle \right).
\]
We complete $\overrightarrow{dt^{\prime}}_{m-1}$\ with $\delta n$ zeros and
reorder the timings so as to obtain $\overrightarrow{dt}_{m-1}$. Then we take
$\overrightarrow{t}_{m}^{\alpha}$ $=\overrightarrow{t}_{m-1}+$ $\alpha$
$\overrightarrow{dt}_{m-1}$. If there exists an $\alpha_{m}$ such that
$F\left(  \overrightarrow{t}_{m}^{\alpha_{m}}\right)  <$ $F\left(
\overrightarrow{t}_{m-1}\right)  $ we set $\overrightarrow{t}_{m}$
$=\overrightarrow{t}_{m}^{\alpha_{m}}$ as our new time-vector, and keep the
same free parameters for the $(m+1)$th step: the permutation
characterizing free-varying timings in the $(m+1)$th step will be the same as
in the $m$th\ step, that is $\sigma_{m}=\sigma_{m-1}$. Otherwise, we take
$\overrightarrow{t}_{m}=\overrightarrow{t}_{m-1}$ as our time-vector and
randomly pick up $MI%
{{}^2}%
$ new free parameters among the $n_{C}$ timings, by choosing a new permutation
$\sigma_{m}$ for the $(m+1)$th step.

\bigskip

We have not said anything about the decoding so far. If the two Hamiltonians $\widehat
{H}_{a}=\widehat
{V}_{a}$ and $\widehat
{H}_{b}=\widehat
{V}_{b}$ can be reversed (note that we assume $\widehat
{H}_{0}=0$), i.e. the sign of $\widehat
{V}_{a}$ and $\widehat{V}_{b}$ can be reversed by altering of the control field 
parameters, the implementation of the decoding
matrix is quite easy: it amounts to reversing $\widehat{V}_{a}$ and
$\widehat{V}_{b}$ and applying the same control timing sequence backwards. 
To be more explicit, one starts by applying $-\widehat{V}_{b}$ during
timing $t_{n_{C}}$, then $-\widehat{V}_{a}$ during $t_{n_{C}-1}$,... , and
finally $-\widehat{V}_{a}$ during $t_{1}$. On the contrary, if $\widehat
{V}_{a}$ and $\widehat{V}_{b}$ cannot be reversed, one cannot apply this technique.
We must use the general non-holonomic control technique, involving $N^{2}$
control parameters, to find timings which realize $\widehat{C}^{-1}$.

The algorithm we have just described was numerically implemented and has
already given satisfying numerical results on a realistic $7$-qubit system
subject to the action of $21$ errors \cite{CPZE}. In the next section, we
deal with another real physical system which lends itself particularly well to
a demonstration of our method.

To conclude this section, let us emphasize that, to our knowledge, there is only a formal link between our method and the so-called ''bang-bang'' control schemes \cite{BB}. Actually, in this kind of techniques, fast and strong pulses are applied which average the interaction Hamiltonian between the system and its environment to zero. By contrast, our method employs pulses which are designed to code information, that is to transfer it into a proper subspace, in which errors act orthogonally : decoding and measurement then allow us to recover initial information. 

\bigskip

\section{Coherence Protection applied to the Rubidium atom}
The goal of this section is to apply our method to a real physical system. As we shall see below, the chosen system, a Rubidium isotope, due to its structure, lends itself particularly well to a straightforward implementation of our technique and allows us to illustrate its different steps quite simply : to be more specific, following the scheme we presented in the previous sections, we show that it is possible to protect one qubit of information encoded on the two spin states of the ground level 5s of
the radioactive isotope $^{78}$Rb against the action of $M=6$ error-inducing
Hamiltonians $\widehat{E}_{m}$. For numerical calculations we considered 3
magnetic Hamiltonians 
\[
\left\{  \widehat{E}_{k}^{\beta}\propto\widehat
{L}_{k}+2\widehat{S}_{k},k=x,y,z\right\},
\] 
and 3 electric Hamiltonians of
second order 
\[
\left\{  \widehat{E}_{k,l}^{\varepsilon}\propto\widehat
{r}_{k}^{2}-\widehat{r}_{l}^{2},k,l=x,y,z,\text{ }k<l\right\}.
\]
In the following, we propose a detailed physical setting which achieves the desired protection operation : in particular, we provide characteristic values of control fields and pulse timings. These different calculated parameters relate to a single isolated atom. As we shall see at the end of this section, when dealing with an ensemble of atoms, serious experimental drawbacks emerge which prevent us from actually implementing our application. Nevertheless, the example considered shows the operationality of our method which is able, in a given physical situation, to provide a precise frame for its implementation.

Before presenting the details of the proposed implementation,
let us motivate the choice of the Rubidium atom. Alkali atoms like Rb are very
interesting for our purpose because of their hydrogen-like behavior. Such an atom 
is the compound of an information subsystem, i.e. the
spin part of the wavefunction, and an ancilla, i.e. the orbital part of the
quantum state. As we shall see, it is easy to increase the dimensionality of
the ancilla by simply pumping the atom towards a shell of higher orbital
angular momentum $L$.

\begin{figure}
[floatfix]
\begin{center}
\includegraphics[
height=3.6002in,
width=2.533in
]%
{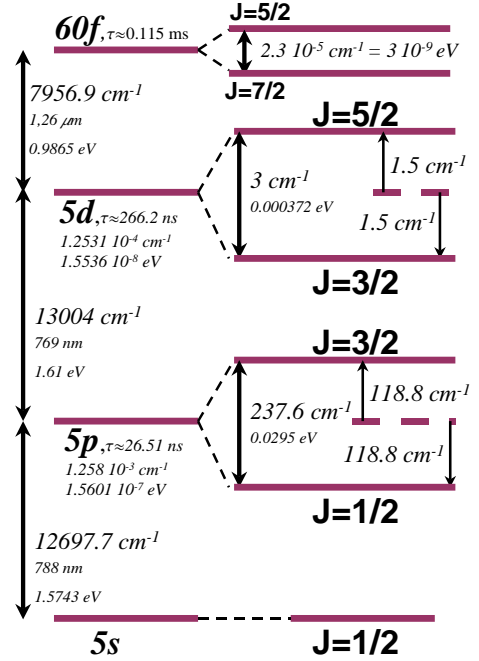}%
\caption{Spectrum of $^{78}$Rb: The useful part of
the spectrum of Rubidium is represented.}%
\label{fig2}%
\end{center}
\end{figure}

We chose $^{78}$Rb among all alkali systems because of its
spectroscopic characterics (Fig.\ref{fig2}) \cite{GALLAGHER,BACHER}. In particular,
$^{78}$Rb has no hyperfine structure (its nuclear spin is $0$) which
ensures that the ground level $5s$ is degenerate: this is necessary for the 
projection scheme as we shall see below. Moreover it has a long enough
lifetime ($\tau\simeq17.66\min$) for the proposed experiment.%

\bigskip

Let us now review each step of our method in detail. As mentioned above, the
information we want to protect is initially encoded on the two spin states
$\left|  \nu_{1}\right\rangle =\left|  5s,j=\frac{1}{2},m_{j}=-\frac{1}%
{2}\right\rangle $ and $\left|  \nu_{2}\right\rangle =\left|  5s,j=\frac{1}%
{2},m_{j}=\frac{1}{2}\right\rangle $ of the ground level $5s$ of the atom:
these two states span the information space $\mathcal{H}_{I}=Span\left[
\left|  \nu_{1}\right\rangle ,\left|  \nu_{2}\right\rangle \right]  $ whose
dimension is in that case $I=2$. The first step of our scheme consists in
adding an ancilla $\mathcal{A}$ to the information system. The role of
$\mathcal{A}$ is played by the orbital part of the wavefunction. In the ground
state ($L=0$), its dimension is $A=2L+1=1$ (roughly speaking, there is no
ancilla). If we want to protect one qubit of information against $M=6$
error-inducing Hamiltonians, we have to increase the dimensionality\ of the
ancilla up to $A=M+1=7$ (Eq.(\ref{Hamming})): this can be achieved by pumping
the atom up to a shell $nf$ ($L=3$). We choose the highly excited Rydberg
state $60f$ so as to make the fine structure as weak as possible (the splitting for $60f$ is approximately $10^{-5}cm^{-1}$ \cite{GALLAGHER}). We shall first consider the fine structure is negligible so that the $N=I\times A=2\times7=14$ basis vectors of the total Hilbert space
$\mathcal{H}=\mathcal{H}_{I}\otimes\mathcal{H}_{A}$ are almost perfectly
degenerate ; the validity of this approximation will be discussed at the end of this section. 
To be more specific, the pumping is done in such a way that
\begin{eqnarray*}
\left|  \nu_{1}\right\rangle & \longrightarrow & \left|  \gamma_{1}%
\right\rangle =\left|  60f,j=\frac{5}{2},m_{j}=-\frac{3}{2}\right\rangle  \\
\left|  \nu_{2}\right\rangle & \longrightarrow & \left|  \gamma_{2}%
\right\rangle =\left|  60f,j=\frac{5}{2},m_{j}=-\frac{1}{2}\right\rangle . 
\end{eqnarray*}
In other words, using the terminology of the previous
sections, the information initially stored in $\mathcal{H}_{I}$ is transferred
into
\begin{widetext}
\[
\mathcal{C}=Span\left[  \left|  \gamma_{1}\right\rangle =\left|
60f,j=\frac{5}{2},m_{j}=-\frac{3}{2}\right\rangle , \right.  \left. \left|  \gamma
_{2}\right\rangle =\left|  60f,j=\frac{5}{2},m_{j}=-\frac{1}{2}\right\rangle
\right]  .
\]
\end{widetext}
The choice of the subspace $\mathcal{C}$ may appear arbitrary at this stage, but it will 
be justified later by the practical feasibility of the
projection process onto $\mathcal{C}$. Let us note that $\mathcal{C}$\ is an
''entangled'' subspace whose basis vectors $\left\{  \left|  \gamma
_{i}\right\rangle \right\}  _{i=1,2}$\ are general entangled states of the
spin and orbital parts: this means (Sec.II) that
the projection step will not consist in a simple measurement of the ancilla
but will involve a more intricate process we shall describe in detail later. 

\begin{figure*}
[floatfix]
\begin{center}
\includegraphics[
height=3.5483in,
width=5.0989in
]%
{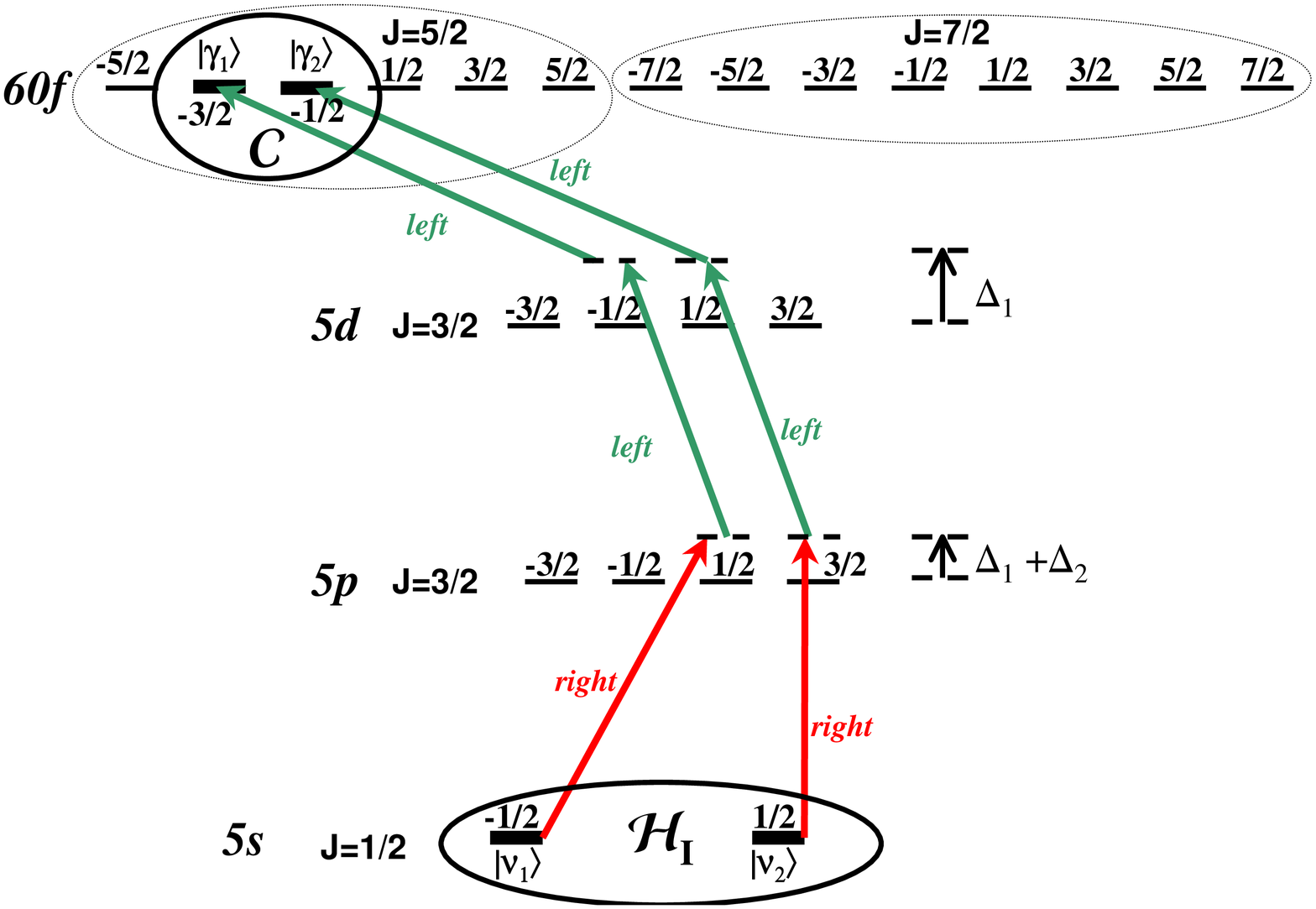}%
\caption{Ancilla adding by Pumping. Photon polarization and
involved sub-Zeeman levels are represented. The fine structure of the Rydberg level $60f$ is not resolvable.}%
\label{fig3}%
\end{center}
\end{figure*}

Practically, the pumping can be achieved as follows. 
One applies three lasers to the atom: the first laser is right polarized
and slightly detuned from the transition $\left(  5s\leftrightarrow5p\right)
$ whereas the second and third lasers are left polarized and slightly detuned
from the transitions $\left(  5p_{\frac{3}{2}}\leftrightarrow5d_{\frac{3}{2}%
}\right)  $\ and $\left(  5d_{\frac{3}{2}}\leftrightarrow60f\right)
$\ respectively. The detunings forbid real one-photon processes: the atom can only 
absorb three photons simultaneously and is thereby excited from the
ground level $5s$ to the Rydberg level $60f$. By using selection
rules, one can construct the allowed paths represented on Fig.\ref{fig3}: these paths 
only couple $\left|  \nu_{1}\right\rangle $ and
$\left|  \nu_{2}\right\rangle $\ to $\left|  \gamma_{1}\right\rangle $ and
$\left|  \gamma_{2}\right\rangle $, respectively.%

\bigskip

\begin{figure*}
[floatfix]
\begin{center}
\includegraphics[
height=3.3088in,
width=4.7556in
]%
{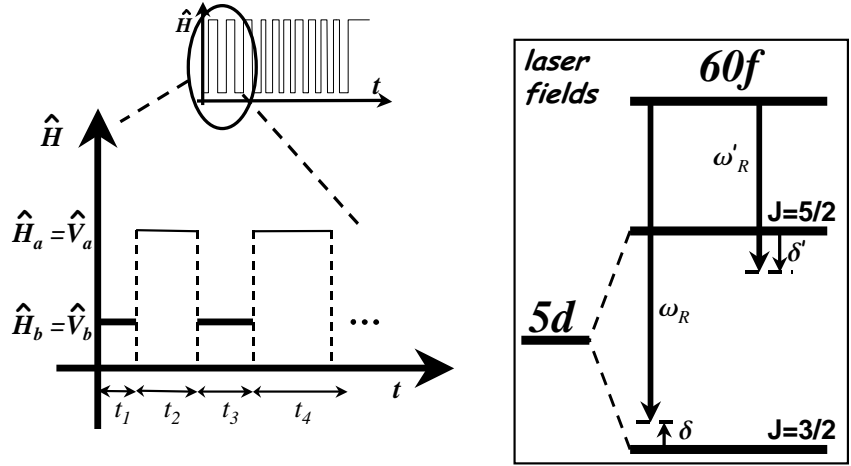}%
\caption{Coding step through the non-holonomic control technique. The two Hamiltonians $\widehat{H}_{a}$\ and $\widehat{H}_{b}$ are alternately applied to the system during pulses of timings
$\left\{  t_{i}(ns)\right\}=$\{3.9763, 6.4748, 4.2274, 3.6259, 2.8717, 3.6281, 7.2263, 6.4260, 4.8070, 5.0394, 6.5242, 4.8890, 4.2400, 7.3834, 4.8653, 5.4799, 4.5341, 4.3099, 6.2959, 3.7346, 6.5293, 6.8586, 6.0749, 5.1213, 4.6806, 3.4985, 3.9909, 4.6701, 4.5168, 6.4702, 4.7787, 5.3476, 3.4567, 3.8009\}. The frequencies of the
laser fields involved in the encoding step are represented on the spectrum of
the Rubidium atom. The fine structure of the Rydberg level $60f$ is not represented.}%
\label{fig4}%
\end{center}
\end{figure*}

The second step consists in encoding the information by the non-holonomic
control technique: to impose the coding matrix on the system, we submit the
atom to $n_{C}=34$ control pulses of timings $\left\{  t_{i}\right\}
_{i=1,...,34}$, during which two different combinations of magnetic and Raman
electric Hamiltonians are alternately applied (see Fig.\ref{fig4}). To be more
explicit, during odd-numbered pulses (''A'' type pulses) we apply a constant
magnetic field
\[
\overrightarrow{B}=\left(
\begin{array}
[c]{c}%
B_{x}=7\;10^{-3}T\\
B_{y}=8.2\;10^{-3}T\\
B_{z}=-6.8\;10^{-3}T
\end{array}
\right)
\]
which is associated with the Zeeman Hamiltonian $\widehat{W}_{Z}$, and two
sinusoidal electric laser fields
\begin{eqnarray*}
\overrightarrow{E}_{a}(t) =\mbox{Re}\left[  \underline
{\overrightarrow{E}}_{a}e^{-i\omega_{R}t}\right],\overrightarrow
{E}_{a}^{\prime}(t)=\mbox{Re}\left[  \underline{\overrightarrow{E}%
}_{a}^{\prime}e^{-i\omega_{R}^{\prime}t}\right], \\
\underline{\overrightarrow{E}}_{a} =\left|
\begin{array}
[c]{c}%
E_{x,a}\\
E_{y,a}e^{-i\varphi_{y,a}}\\
0
\end{array}
\right. ,\underline{\overrightarrow{E}}_{a}^{\prime}=\left|
\begin{array}
[c]{c}%
E_{x,a}^{\prime}\\
E_{y,a}^{\prime}e^{-i\varphi_{y,a}^{\prime}}\\
0
\end{array}
\right.,
\end{eqnarray*}
whose frequencies $\omega_{R}$ and $\omega_{R}^{\prime}$\ are respectively
slightly detuned from the two transitions $\left(  60f\leftrightarrow
5d,j=\frac{3}{2}\right)  $ and $\left(  60f\leftrightarrow5d,j=\frac{5}%
{2}\right)  $ (detunings $\delta$ and $\delta^{\prime}$). The characteristic
values of these fields are
\begin{eqnarray*}
E_{x,a} & = & E_{x,a}^{\prime}=8.5\;10^{5}V.m^{-1}  \\
E_{y,a} & = & E_{y,a}^{\prime}=5.2\;10^{6}V.m^{-1}  \\
\varphi_{y,a} & = & \varphi_{y,a}^{\prime}=2.3  \\
\hbar\omega_{R} & = & 0.986324\;eV=7955.14\;cm^{-1}  \\
\delta & = & -0.000010\;eV=-0.080654\;cm^{-1} \\
\hbar\omega_{R}^{\prime} & = & 0.986676\;eV=7958.14\;cm^{-1}  \\
\delta^{\prime} & = & 0.000010\;eV=0.080654\;cm^{-1}. 
\end{eqnarray*}
The intensity of the laser beams are typically of the order of $2$ $10^{8}W.cm^{-1}$.
The Raman Hamiltonian associated with these fields is denoted by $\widehat
{W}_{R,A}$. The total perturbation is $\widehat{V}_{a}=\widehat{W}%
_{Z}+\widehat{W}_{R,A}$. During even-numbered pulses (''B'' type pulses), we
apply the same magnetic field as for A type pulses, which is experimentally
convenient, and two sinusoidal electric laser fields
\begin{eqnarray*}
\overrightarrow{E}_{b}(t)=\mbox{Re}\left[  \underline
{\overrightarrow{E}}_{b}e^{-i\omega_{R}t}\right]  ,\overrightarrow
{E}_{b}(t)=\mbox{Re}\left[  \underline{\overrightarrow{E}}_{b}%
^{\prime}e^{-i\omega_{R}^{\prime}t}\right], \\
\mbox{where }\underline{\overrightarrow{E}}_{b}=\left|
\begin{array}
[c]{c}%
E_{x,b}\\
E_{y,b}e^{-i\varphi_{y,b}}\\
0
\end{array}
\right. ,\underline{\overrightarrow{E}}_{b}^{\prime}=\left|
\begin{array}
[c]{c}%
E_{x,b}^{\prime} \\
E_{y,b}^{\prime}e^{-i\varphi_{y,b}^{\prime}}  \\
0
\end{array}
\right. ,
\end{eqnarray*}
whose frequencies are the same as above and whose characteristics values are
\begin{eqnarray*}
E_{x,b} & = & E_{x,b}^{\prime}=-5.2\text{ }10^{6}V.m^{-1} \\
E_{y,b} & = & E_{y,b}^{\prime}=8.5\;10^{5}V.m^{-1} \\
\varphi_{y,a} & = & \varphi_{y,a}^{\prime}=2.3.  
\end{eqnarray*}
The Raman Hamiltonian associated with these fields is denoted by $\widehat
{W}_{R,B}$. The corresponding perturbation is $\widehat{V}_{b}=\widehat{W}%
_{Z}+\widehat{W}_{R,B}$. Therefore, since the fine structure of the level $60f$ is neglected, the unperturbed Hamiltonian
$\widehat{H}_{0}$ is taken to be $0$ and the total Hamiltonian has the form:
$\widehat{H}_{A}=\widehat{V}_{a}$ during ''A'' pulses, $\widehat{H}_{B}=\widehat{V}_{b}$ during ''B'' pulses.
The $34$ different timings have been calculated so that
\[
\widehat{U}(t_{1},...,t_{34})=e^{-i\widehat{H}_{B}t_{n_{C}}}
e^{-i\widehat{H}_{A}t_{n_{C}-1}}\ldots e^{-i\widehat{H}_{A}t_{1}}=\widehat{C}%
\]
meets conditions (\ref{matcod}). At the end of the coding step the
information is transferred into the code space $\widetilde{\mathcal{C}%
}=\widehat{C}\mathcal{C}$, encoded on the codewords $\left\{  \left|
\widetilde{\gamma}_{i}\right\rangle =\widehat{C}\left|  \gamma_{i}%
\right\rangle \right\}  _{i=1,2}$.
\qquad

As can be easily checked from Fig.\ref{fig4} the total duration of a control
period ($\simeq125ns$) is approximately $10^{3}$ times shorter than the
lifetime of $60f$ Rydberg state which is approximately $0.115ms$ as can be
calculated from \cite{GALLAGHER}. The different pulse timings range between 
$2.9ns$ and $7.4ns$, which are feasible.

\bigskip

\begin{figure*}
[ptb]
\begin{center}
\includegraphics[
height=3.6313in,
width=5.2217in
]%
{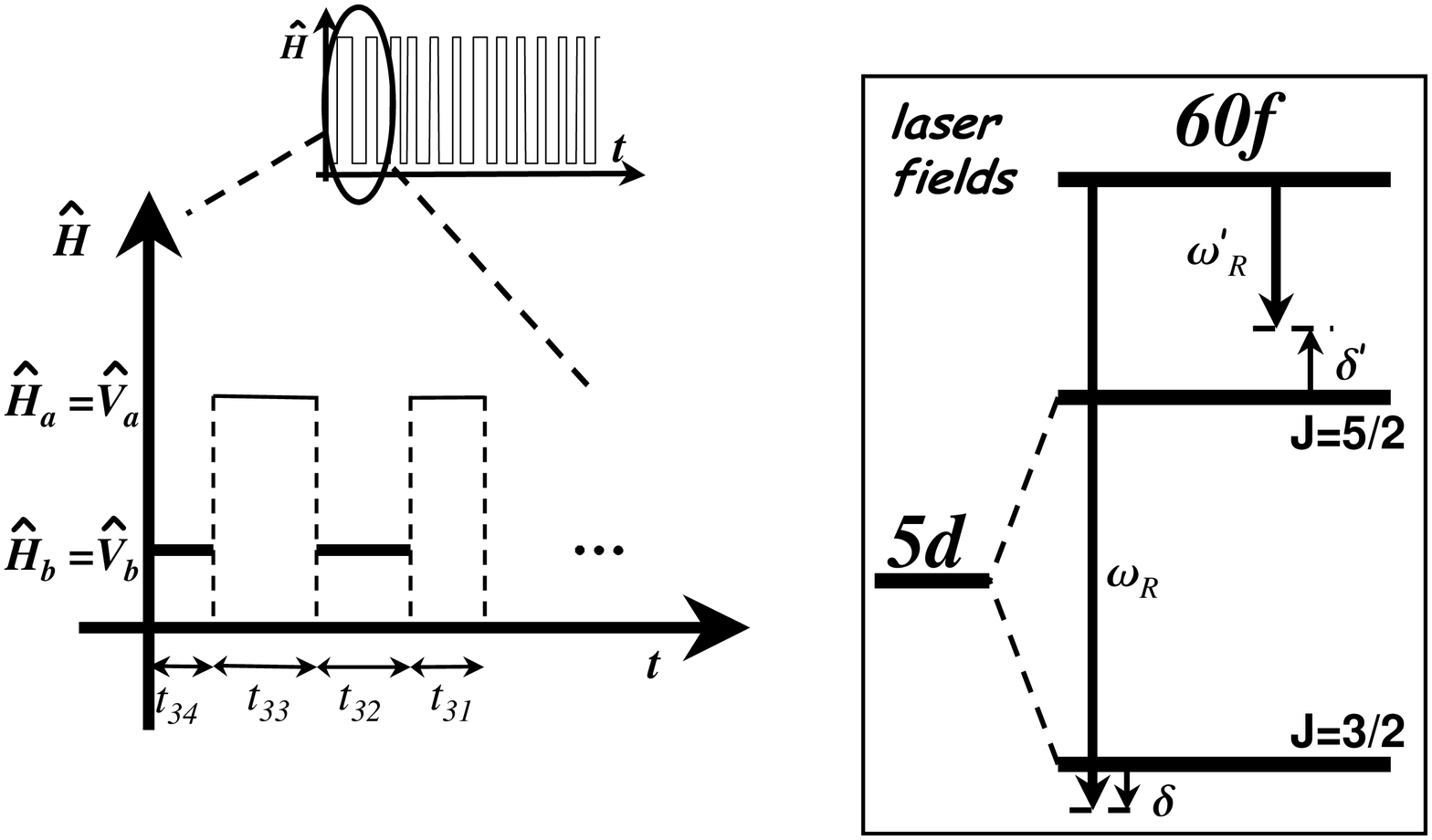}%
\caption{Decoding step by the non-holonomic control technique.
We reverse the magnetic field and the detunings of electric fields, as
represented on the spectrum of the Rubidium atom, and apply the same control
sequence as for coding (same timings) backwards. The fine structure of the level $60f$ is not represented.}%
\label{fig5}%
\end{center}
\end{figure*}

After a short time, the information stored in the system acquires a small
erroneous component due to the action of the error Hamiltonians, which is
orthogonal to the code space $\widetilde{\mathcal{C}}$. Then, we apply the
decoding matrix $\widehat{C}^{-1}$ to the atom as suggested at the end of Sec.III. We
reverse $\overrightarrow{B}$ and the detunings $\delta$ and $\delta^{\prime}$,
and leave all the other values unchanged (this amounts to taking the
opposite of Hamiltonians $\widehat{H}_{A}$ and $\widehat{H}_{B}$), and apply the same 
sequence of control pulses backwards: we start with
an ''A'' pulse whose timing is $t_{n_{C}}$, then apply a ''B'' pulse during
$t_{n_{C}-1},$ etc. (see Fig.\ref{fig5}). The decoding step yields an
erroneous state whose projection onto $\mathcal{C}$ is the initial information state.%

\begin{figure*}
[ptb]
\begin{center}
\includegraphics[
height=3.544in,
width=5.1162in
]%
{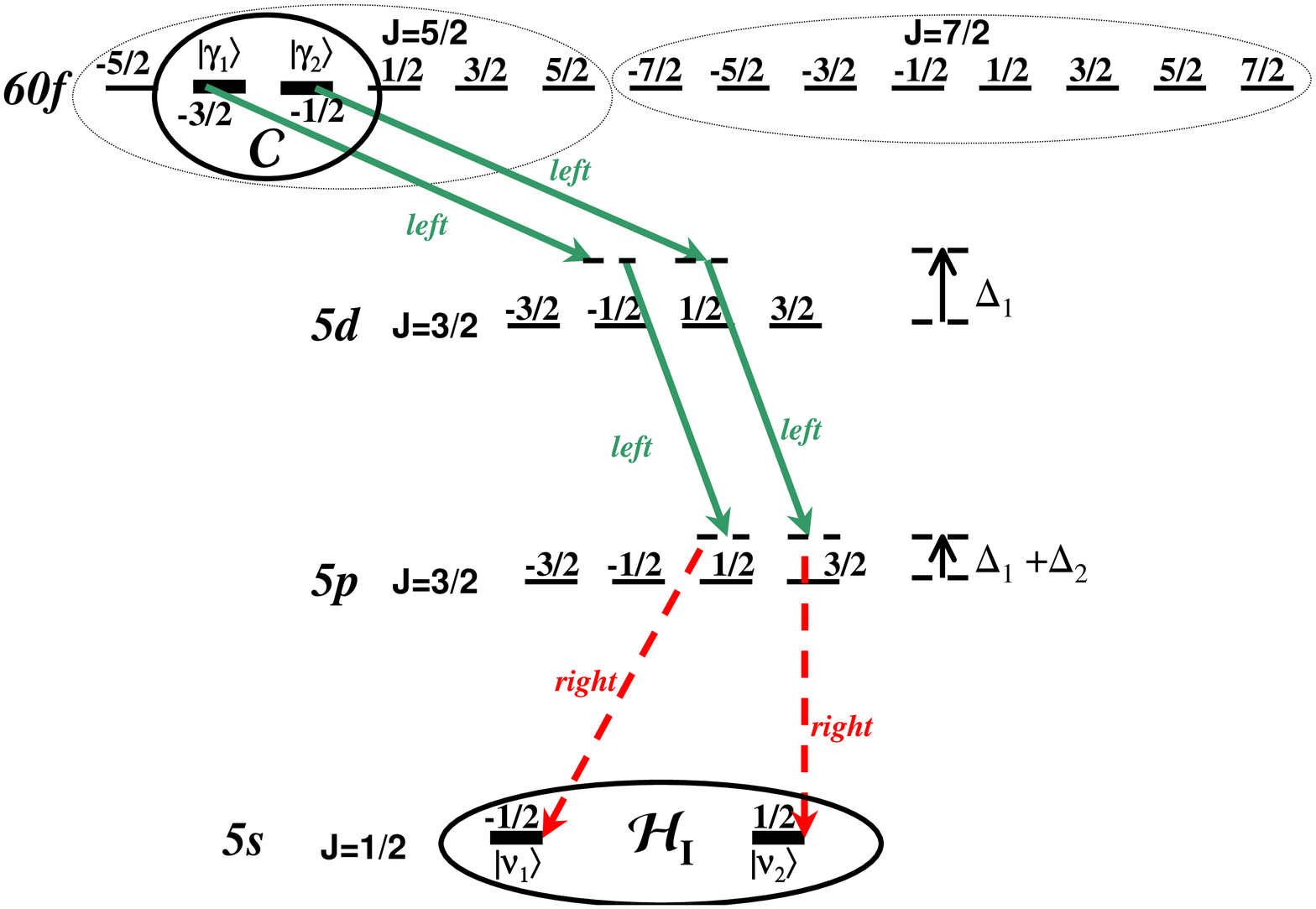}%
\caption{Projection path. The lasers involved are marked by solid arrows, 
the spontaneous photon is represented by a dashed arrow. The
different polarizations are specified. The fine structure of the level $60f$ is not represented.}%
\label{fig6}%
\end{center}
\end{figure*}

In the last step the erroneous state vector is projected onto the subspace
$\mathcal{C}$ to recover the initial information. Projection is a non-unitary
process which cannot be achieved through a Hamiltonian process, but requires
the introduction of irreversibility. To this end, we make use of a path which
is symmetric with the pumping step, and consists in two stimulated and one
spontaneous emissions. To be more explicit, we apply two left circularly
polarized lasers (see Fig.\ref{fig6}) slightly detuned from the transitions
$\left(  60f\longleftrightarrow5d,j=\frac{3}{2}\right)  $ and $\left(
5d,j=\frac{3}{2}\longleftrightarrow5p,j=\frac{3}{2}\right)$. Due to these laser
fields, the atom is likely to fall towards the ground state and emit two
stimulated and one spontaneous photons.

Using the selection rules, one can infer that, if a circularly right-polarized
spontaneous photon is emitted, the only states to be coupled to the ground
level are $\left|  \gamma_{1}\right\rangle $ and $\left|  \gamma
_{2}\right\rangle $ to $\left|  \nu_{1}\right\rangle $ and
$\left|  \nu_{2}\right\rangle$, respectively (see Fig.\ref{fig6}). This means that the
emission of a right polarized spontaneous photon brings the ''correct'' part
of the state vector back into $\mathcal{H}_{I}=Span\left[  \left|  \nu
_{1}\right\rangle ,\left|  \nu_{2}\right\rangle \right]  $. On the contrary,
the other cases - ''left polarized'', ''linearly polarized spontaneous
photon'', or ''no photon at all''- do not lead to the right projection
process.%

The ''left-polarized photon'' and ''no photon emitted'' cases are quite
unlikely: indeed the probability that they occur is proportional to the
square of the error amplitude, that is to the square of the Zeno interval $T$,
which is very short. The ''linearly polarized photon'' case
is quite annoying because it mixes the two paths $\left|  \gamma
_{1}\right\rangle \longrightarrow\left|  \nu_{1}\right\rangle $ and $\left|
\gamma_{2}\right\rangle \longrightarrow\left|  \nu_{2}\right\rangle $. This
parasitic process and its relative probability must be suppressed, with
respect to the process followed by the ''right--polarized'' photon emission.
This can be done by launching the $^{78}$Rb atom, previously cooled, into a Fabry-Perot cavity, in an atomic fountain manner (fine tuning of the lasers driving the $60f~-~5d$ and $5d~-~5p$ transition will be
necessary to avoid reflection of the external laser radiation from the
cavity). The decay rate for the 3-photon transition $\left|  \gamma
_{i}\right\rangle \longrightarrow\left|  \nu_{i}\right\rangle $ is
\begin{widetext}
\begin{displaymath}
\Gamma_{\gamma_{i}\nu_{i}}=2\pi\left|  \frac{d_{\gamma_{i}\lambda_{j}}E_{1}%
}{\hbar\Delta_{1}}\right|  ^{2}\left|  \frac{d_{\lambda_{j}\mu_{k}}E_{2}%
}{\hbar(\Delta_{1}+\Delta_{2})}\right|  ^{2}\overline{2\pi\hbar ck_{s}\left|
\overrightarrow{d}_{\mu_{k}\nu_{i}}\overrightarrow{e}_{R}^{\ast}\right|
^{2}\varrho\left(  \overrightarrow{k}_{s}\right)  },
\end{displaymath}
\end{widetext}
where $\overrightarrow{k}_{s}$ is the wave vector of the spontaneously emitted
photon, $\overrightarrow{e}_{R}$ is the left-polarized photon polarization
unit vector, $\varrho\left(  \overrightarrow{k}_{s}\right)  $ is the density
of states (normalized to the cavity volume) for the cavity field at
$\overrightarrow{k}_{s}$, and the bar denotes averaging over the directions of
$\overrightarrow{k}_{s}$. The transition dipole moments are denoted by
$d_{ab}$: during the projective process the states coupled to $\left|
\gamma_{1}\right\rangle $ and $\left|  \gamma_{2}\right\rangle $ are,
respectively, 
\begin{widetext}
\begin{eqnarray*}
\left\{  \left|  \lambda_{1}\right\rangle =\left|  5d,j=\frac
{3}{2},m_{j}=-\frac{1}{2}\right\rangle , \left|  \lambda_{2}\right\rangle
=\left|  5p,j=\frac{3}{2},m_{j}=+\frac{1}{2}\right\rangle \right\}, \\ \left\{  \left|  \mu_{1}\right\rangle =\left|  5d,j=\frac{3}{2},m_{j}%
=\frac{1}{2}\right\rangle , \left|  \mu_{2}\right\rangle =\left|  5p,j=\frac
{3}{2},m_{j}=\frac{3}{2}\right\rangle \right\}.
\end{eqnarray*} 
\end{widetext}

The enhancement (by the
presence of cavity) of the density of states for the modes propagating
paraxially to the $z$-axis ensures that
\[
\Gamma_{\gamma_{1}\nu_{1}},\Gamma_{\gamma_{2}\nu_{2}}\gg\left|  \frac
{d_{\gamma_{i}\lambda_{j}}E_{1}}{\hbar\Delta_{1}}\right|  ^{2}\left|
\frac{d_{\lambda_{j}\mu_{k}}E_{2}}{\hbar(\Delta_{1}+\Delta_{2})}\right|
^{2}\gamma,
\]
where $\gamma$ is the decay rate of $\left|  5p,j=\frac{3}{2}%
,m_{j}=+\frac{1}{2}\right\rangle $ into $\left|  5s,j=\frac{1}{2},m_{j}%
=+\frac{1}{2}\right\rangle $, so that the undesired process followed by the
$\pi$-photon emission is relatively less important than it were in free
space. For the density matrix elements $\rho_{ab}$ the following
system of equations can be written ($i=0,1$):
\begin{eqnarray}
\dot{\rho}_{\gamma_{i}\gamma_{i}}=-\Gamma_{\gamma_{i}\nu_{i}}\dot{\rho
}_{\gamma_{i}\gamma_{i}}, \nonumber \\
\dot{\rho}_{\nu_{i}\nu_{i}}=\Gamma_{\gamma_{i}\nu_{i}}\dot{\rho}%
_{\gamma_{i}\gamma_{i}}, \nonumber \\
\dot{\rho}_{\gamma_{1}\gamma_{2}}=-\frac{1}{2}(\Gamma_{\gamma_{1}\nu_{1}%
}+\Gamma_{\gamma_{2}\nu_{2}})\rho_{\gamma_{1}\gamma_{2}}, \nonumber \\
\dot{\rho}_{\nu_{1}\nu_{2}}=\sqrt{\Gamma_{\gamma_{1}\nu_{1}}%
\Gamma_{\gamma_{2}\nu_{2}}}\rho_{\gamma_{1}\gamma_{2}}. \nonumber 
\end{eqnarray}
To avoid dephasing which would corrupt the information, the
coherence matrix element $\rho_{\gamma_{1}\gamma_{2}}$ must be transferred
with the maximum efficiency into $\rho_{\nu_{1}\nu_{2}}$: the efficiency
\[
\eta=\frac{2\sqrt{\Gamma_{\gamma_{1}\nu_{1}}\Gamma_{\gamma_{2}\nu_{2}}}%
}{\Gamma_{\gamma_{1}\nu_{1}}+\Gamma_{\gamma_{2}\nu_{2}}}%
\]
is thus crucial. According to the Wigner-Eckart theorem,
\[
\frac{\Gamma_{\gamma_{1}\nu_{1}}}{\Gamma_{\gamma_{2}\nu_{2}}}=\left(
\frac{C_{3/2~-1/2~1~-1}^{5/2~-3/2}C_{3/2~1/2~1~-1}^{3/2~-1/2}C_{1/2~-1/2~1~1}%
^{3/2~1/2}}{C_{3/2~1/2~1~-1}^{5/2~-1/2}C_{3/2~3/2~1~-1}^{3/2~1/2}%
C_{1/2~1/2~1~1}^{3/2~3/2}}\right)  ^{2},
\]
where in the right-hand-side the ratio of the products of the Clebsch-Gordan
coefficients corresponding to the transitions stands. These coefficients, which can
be found in \cite{Varshalovich}, lead to $\eta=12\sqrt
{2}/17\approx0.99827$. In other words, the probability of error during the Zeno
projection stage due to the small difference of the Clebsch-Gordan coefficient
products for the two paths is equal to or less than $1-\eta\approx0.00173$
(the equality is reached if the initial state is $(\left|  0\right\rangle
\pm\left|  1\right\rangle )/\sqrt{2}$). Note that the states $60f$,
$5d$, and $5p$ have finite lifetimes $\tau_{k}$ (see Fig.\ref{fig2}). Thus the
transition rates $\Gamma_{\gamma_{i}\nu_{i}}$ must be much larger than
$1/\tau_{60f}$, $\left|  \frac{d_{\gamma_{i}\lambda_{j}}E_{1}}{\hbar\Delta
_{1}}\right|  ^{2}/\tau_{5d}$, and $\left|  \frac{d_{\gamma_{i}\lambda_{j}%
}E_{1}}{\hbar\Delta_{1}}\right|  ^{2}\left|  \frac{d_{\lambda_{j}\mu_{k}}%
E_{2}}{\hbar(\Delta_{1}+\Delta_{2})}\right|  ^{2}/\tau_{5p}$, in order to
diminish errors caused by the decay of these unstable states.

To complete the projection step, one has to transfer the atom in its coherent
superposition back to the $60f$ state: this is achieved by the same pumping
sequence as in the first step. The mismatch of the Clebsch-Gordan coefficient
products will cause again the error probability $1-\eta$. The information is
then restored with very high probability and the system is ready to undergo a
new protection cycle.

\bigskip

From the beginning of this section we have neglected the fine structure
splitting of the level $60f$, which is approximately $2.10^{-5}cm^{-1}$ and
corresponds to a period $\tau_{f}\sim1.5\mu s$. To conclude this section, let
us now take it into account and see its effect on each step of our scheme.

Obviously the pumping and projection steps will not be affected by the fine
structure, since the information-carrying vectors $\left\{  \left|
\gamma_{1}\right\rangle ,\left|  \gamma_{2}\right\rangle \right\}  $ belong to
the same multiplet $\left(  J=5/2\right)  $.

The coding and decoding steps are neither modified by the existence of the
fine structure. Indeed, since the typical period of the fine structure
Hamiltonian $\tau_{f}\sim1.5\mu s$\ is more than $10$ times longer than the
total duration of the coding or decoding steps, it is legitimate to neglect
its effect.

The influence of the fine structure on the free evolution period during which
errors are likely to occur is more complicated to study in the general case.
Yet, two simple limiting regimes can be considered. If the spectrum of the
coupling functions $f_{m}(t)$'s is very narrow (\emph{i.e.} if the variation
timescale of the $f_{m}(t)$'s is much longer than $\tau_{f}$), one can show
that our scheme applies directly as though there were no fine structure,
provided the error Hamiltonians $\left\{  \widehat{E}_{m}\right\}  $ are
replaced by $\left\{  \widehat{E}_{m}^{(0)}\right\}  $, where $\widehat{E}%
_{m}^{(0)}$ is obtained from $\widehat{E}_{m}$\ by simply setting to zero the
rectangular submatrices which couple the two multiplets $\left(
J=5/2,7/2\right)  $. The second limiting regime corresponds to a very broad
spectrum for the $f_{m}(t)$'s (variation timescale much shorter than $\tau
_{f}$): in that case, one can show that our scheme applies provided one
chooses a Zeno interval $T$ multiple of $\tau_{f}$.

In all this section, we implicitly supposed that the Rubdium atom was alone ; but in actual experiments, one usually works with an ensemble of atoms : this generates serious experimental drawbacks which we deal with now. Rydberg atoms are sensitive to Doppler effect : nevertheless, in the case of cold atoms, this is negligible. But the most dramatic effect is due to interactions between atoms such as dipolar forces \cite{PRSG02}. In a standard Magneto Optical Trap containing about $1000$ atoms in  Rydberg states $(n\simeq 60)$, the typical energy of these interactions is 1 MHz, corresponding to a dephasing time of $1ms$ \cite{SRLAMW04}. As different atoms see different environments, and are thus subject to different interactions, it will be impossible to properly code and thus protect the information stored in the different atoms of the ensemble. Beyond these problems, we nevertheless want to emphasize the demonstrative value of our example : the system considered here (Rubidium isotope in a Rydberg state), though not completely satisfactory from an experimental point of view, is indeed quite practical for a straightforward demonstration of our scheme, since information carrying subsystem and ancilla are clearly identified, and every step is "simply" achieved. Other systems must be found, which will be addressed in future publications ; however, the application considered here has already suggested the physical relevance and applicability of our method.

\section{Conclusions}

In this paper, an original scheme has been presented which allows to protect
the quantum coherence stored in a information system $\mathcal{I}$ against the
action of a set of $M$ given error-inducing Hamiltonians $\widehat{E}_{k} $.

The information initially stored in the Hilbert space $\mathcal{H}%
_{I}$ of the information system is transferred into a subspace $\mathcal{C}$
of the Hilbert space $\mathcal{H}=\mathcal{H}_{I}\otimes\mathcal{H}_{A}$ of
the compound system $\mathcal{I}\otimes\mathcal{A}$ formed through adding an
auxiliary system $\mathcal{A}$\ called ancilla to the main system. A
multidimensional generalization of the QZE has been presented
which makes it possible to protect such a subspace against the action of the
$\widehat{E}_{k}$'s, provided the dimension $A$ of the ancilla meets the
Hamming bound $A\geq M+1$. The information is thus encoded in another
subspace $\widetilde{\mathcal{C}}$, called the ''code space'', through the
application of the coding matrix $\widehat{C}$: in this appropriate subspace,
the error-inducing Hamiltonians $\widehat{E}_{k}$ act orthogonally. After a
short time, the information thus contains a small orthogonal erroneous
component due to the action of the $\widehat{E}_{k}$ ; it is then decoded by
application of $\widehat{C}^{-1}$ and restored by an appropriate physical
measurement which projects the state vector onto $\mathcal{C}$ with very high
probability. The repetition of this sequence as long as needed protects the
information stored in the system.

A physical achievement of the coding and decoding steps have been proposed
which employs the non-holonomic control technique. The different algorithmic tools
needed to implement our scheme have been presented.

Finally, an application has been proposed which makes use of the Rubidium
atom. One qubit of information is encoded in the spin states of the atom
whereas the orbital part plays the role of the ancilla. A realistic physical
setting has been proposed: in particular, a projection process based on the
spontaneous emission has been suggested.

\acknowledgments{E.B. thanks Annik Bachelier (laboratoire Aim\'{e} Cotton) for her help. The support of EU(QUACS RTN) and the computational resources of IDRIS-CNRS, Orsay, are kindly acknowledeged. 
I.D. was supported by NSF grant CCR-0097125. I.M. thanks the program Russia Leading Scientific Schools (grant 1115.2003.2) for support.}

\appendix*

\section{Explicit derivation of the code subspace}

In this appendix we deal with a particular physical situation in which the
code subspace $\widetilde{\mathcal{C}}$ can be explicitly derived. We consider
an atom with zero nuclear spin on the level characterized by the orbital angular momentum 
$L$. The electronic spin of the atom is $S=1/2$. The natural basis
wave functions are $\left|  L,M_{L};S,M_{S}\right\rangle $. A qubit of
information is encoded on the two states
\begin{eqnarray*}
\left|  \widetilde{\gamma}_{i}\right\rangle & = & \left|  JLSM_{J_{i}}\right\rangle
\\ & = & \sum_{M_{L},M_{S}}C_{LM_{L}SM_{S}}^{JM_{J_{i}}}\left|  L,M_{L};S,M_{S}%
\right\rangle ,\quad i=1,2,
\end{eqnarray*}
where $C_{LM_{L}SM_{S}}^{JM_{J}}$ is the Clebsch-Gordan coefficient. In the
scheme we proposed for a Rubidium atom, $L=3$, $J=5/2$,
$M_{J_{1}}=-3/2$, $M_{J_{2}}=-1/2$, that is 
\begin{eqnarray*}
\left|  \gamma_{1}\right\rangle
=\left|  60f,j=\frac{5}{2},m_{j}=-\frac{3}{2}\right\rangle \\ 
\left|
\gamma_{2}\right\rangle =\left|  60f,j=\frac{5}{2},m_{j}=-\frac{1}%
{2}\right\rangle.
\end{eqnarray*}
We want to protect this information against the action of
6 independent error-inducing Hamiltonians $\widehat{E}_{k}$, 3 magnetic and 3
electric interaction Hamiltonians. We shall see that the code space
$\widetilde{\mathcal{C}}=Span\left[  \left|  \widetilde{\gamma}_{1}%
\right\rangle ,\left|  \widetilde{\gamma}_{2}\right\rangle \right]  $ can be
simply built from physical considerations on the action of the Hamiltonians
$\widehat{E}_{k}$.

Let us first consider magnetic errors. The interaction Hamiltonian of the atom
with the magnetic field $\overrightarrow{B}$ directed along the \textit{k}-th
axis ($k=x,y,z$) is
\[
\widehat{E}_{k}^{\beta}=\mu_{B}B_{k}(\widehat{L}_{k}+2\widehat{S}_{k}),
\]
$\mu_{B}$ being the Bohr magneton. Remembering that $\widehat{L}_{x}%
=\frac{\widehat{L}_{+}+\widehat{L}_{-}}{2}$, $\widehat{L}_{y}=\frac
{\widehat{L}_{+}-\widehat{L}_{-}}{2i}$, where $\widehat{L}_{+}$ ($\widehat
{L}_{-}$) is the operator increasing (lowering) the $z$-projection of the
orbital angular momentum, and the similar relations for the spin operators, one can
conclude that a pair of the states with definite $z$-projections of orbital
and spin angular momenta is a good basis for the code subspace if the difference of
the of their quantum number $M_{L}\equiv L_{z}$ is greater than or equal to 2.
To use this option, one needs to consider the error caused by a
magnetic field oriented along $z$. The states with definite $L_{z},\,S_{z}$
are the eigenstates of the Hamiltonian $\widehat{E}_{k}^{\beta}$. A general
superposition of two such states must not be rotated in the Hilbert space
under the action of $\widehat{E}_{z}^{\beta}$. This means that the eigenvalues
must be equal to each other. Thus the states
\begin{eqnarray}
\left|  \widetilde{\gamma}_{1}^{\beta}\right\rangle & = & \left|
L,M_{L};S,M_{S}=+1/2\right\rangle , \nonumber \\
\left|  \widetilde{\gamma}_{2}^{\beta}\right\rangle & = & \left|
L,M_{L}+2;S,M_{S}=-1/2\right\rangle , \nonumber 
\end{eqnarray}
with $M\leq L-2$. constitute a good code basis for protecting one qubit
against the action of the $\widehat{E}_{k}^{\beta}$.

We shall now consider errors caused by quasistatic electric fields. The static
Stark shift of a level with zero fine splitting, i.e., a highly excited
Rydberg level, like $60f$ in Rubidium, caused by the electric field $\overrightarrow
{\mathcal{E}}$ oriented along $z$, is given by $const-b\mathcal{E}^{2}M_{L}%
^{2}$. The value $b$ characterizes the polarizability of the atom in the given
state. Omitting the irrelevant constant, we may represent the Hamiltonian of
the atom-field interaction (with respect to the particular manifold of
sublevels of the given atomic state) by the operator
\begin{equation}
\widehat{E}_{k}^{\varepsilon}=-b\mathcal{E}_{k}^{2}\widehat{L}_{k}^{2},\quad
k=x,y,z. \label{he}%
\end{equation}
Note, that since the fine splitting is zero, the spin variables are unaffected
by the Stark effect. Rewriting the operators $\widehat{E}_{x,y}^{\varepsilon}$
in terms of $\widehat{L}_{z}$,
\begin{eqnarray}
\widehat{E}_{x}^{\varepsilon} & = & -b\mathcal{E}_{x}^{2}\left[  \frac{1}%
{4}\widehat{L}_{+}^{2}+\frac{1}{4}\widehat{L}_{-}^{2}+\frac{1}{2}%
L(L+1)-\frac{1}{2}\widehat{L}_{z}^{2}\right]  , \nonumber \\
\widehat{E}_{y}^{\varepsilon} & = & -b\mathcal{E}_{y}^{2}\left[  -\frac{1}%
{4}\widehat{L}_{+}^{2}-\frac{1}{4}\widehat{L}_{-}^{2}+\frac{1}{2}%
L(L+1)-\frac{1}{2}\widehat{L}_{z}^{2}\right]  , \nonumber 
\end{eqnarray}
one can conclude that the basis of the coding space can be a pair of
states of opposite $S_{z}$ and opposite $L_{z}$ (so that $|L_{z}|$ is the
same for both of these states). Indeed, these states are not mixed by the
Hamiltonian ({\ref{he})), which does not cause spin flips. The error vector
is always orthogonal to any their superposition, as can be easily seen. Among
various code subspaces protecting against electric errors there is one that
protects against magnetic errors, too. The basis vectors of this subspace are
\begin{eqnarray}
\left|  \widetilde{\gamma}_{1}^{\beta,\varepsilon}\right\rangle & = & \left|
L,M_{L}=-1;S,M_{S}=+1/2\right\rangle , \nonumber \\
\left|  \widetilde{\gamma}_{2}^{\beta,\varepsilon}\right\rangle & = & \left|
L,M_{L}=+1;S,M_{S}=-1/2\right\rangle . \nonumber 
\end{eqnarray}
}

It may happen that for singlet electronic states of atoms with non-zero
nuclear spin, whose nuclear magnetic moment $\mu_{nucl}$ is incommensurate
with $\mu_{B}$, one cannot apply this explicit derivation of the code space
for the correction of \textit{both} the ``electric'' and ``magnetic'' errors.
One has then to look for a more complex coding transformation through the
algorithm we presented in Sec.III.

\end{document}